\documentclass[amsmath,11pt,amssymb,preprint,nofootinbib]{revtex4}
\usepackage{amsfonts} 
\usepackage{graphicx} 
\usepackage{epsfig}
\usepackage{bm}

\newcommand{\diracslash}[1]{#1\llap{/\kern2pt}}

\newcommand{\be}{\begin{equation}}
\newcommand{\ee}{\end{equation}}
\newcommand{\bea}{\begin{eqnarray}}
\newcommand{\eea}{\end{eqnarray}}
\newcommand{\ba}[1]{\begin{array}{#1}}
\newcommand{\ea}{\end{array}}

\newcommand{\bt}{\begin{tabular}}
\newcommand{\et}{\end{tabular}}

\newcommand{\beas}{\begin{eqnarray*}}
\newcommand{\eeas}{\end{eqnarray*}}

\begin{document}

\title{Upsilon decay widths in magnetized asymmetric nuclear matter}

\author{Amruta Mishra}
\email{amruta@physics.iitd.ac.in}
\affiliation{Department of Physics, Indian Institute of Technology, Delhi,
Hauz Khas, New Delhi -- 110 016, India}

\author{S.P. Misra}
\email{misrasibaprasad@gmail.com}
\address{Institute of Physics, Bhubaneswar -- 751005, India} 


\begin{abstract}
The in-medium partial decay widths of $\Upsilon (4S) \rightarrow B\bar B$
in magnetized asymmetric nuclear matter are studied using a
field theoretic model for composite hadrons with quark (and antiquark)
constituents. $\Upsilon (4S)$ is the lowest bottomonium 
state which can decay to $B\bar B$ in vacuum.
The medium modifications of the decay widths 
of $\Upsilon (4S)$ to $B\bar B$ pair in magnetized matter
arise due to the mass modifications of the decaying $\Upsilon (4S)$ 
as well as of the produced $B$ and $\bar B$ mesons. 
The in-medium masses of the open bottom meson in magnetized
nuclear matter are computed from their interactions with 
the nucleons and the scalar mesons within a chiral effective model.
The mass modification of the $\Upsilon(4S)$ arises due to the
medium modification of a scalar dilaton field, which is introduced
in the model to simulate the gluon condensates of QCD.
The effects of the anomalous magnetic moments
for the proton and neutron are taken into consideration
in the present investigation.
The presence of the external magnetic field
is observed to lead to different mass modifications within the
$B (B^+, B^0)$ as well as the $\bar B (B^-, \bar {B^0})$ doublets,
even in isospin symmetric nuclear matter. This is due to the 
difference in the interactions of the proton and the neutron 
to the electromagnetic field. The charged 
$B^{\pm}$ mesons have additional contributions from the
Landau energy levels, leading to positive shifts in their
masses in the presence of a magnetic field. 
In the presence of an external magnetic field, 
there are contributions to the masses of the $B$, $\bar B$ mesons 
and $\Upsilon(4S)$ state (longitudinal component)
due to the pseudoscalar meson-vector meson (PV)
mixing ($B-B^*$, $\bar B- \bar {B^*}$
and $\Upsilon(4S)-\eta_b(4S)$ mixings), 
which are also considered in the present study. 
The PV mixing effects are observed to be the 
dominant contributions to the mass shifts of 
these mesons, which lead to appreciable modifications
in the decay widths of $\Upsilon (4S)$ to the neutral
($B^0 \bar {B^0}$) and the charged ($B^+ B^-$) pairs
in the presence of a magnetic field. These should have 
observable consequence in the production of open bottom mesons
and bottomonium states at LHC and RHIC, where huge magnetic fields
are produced in ultra-relativistic peripheral heavy ion
collisions. 

\end{abstract}

\maketitle

\def\bfm#1{\mbox{\boldmath $#1$}}
\def\bfs#1{\mbox{\bf #1}}

\section{Introduction}

The study of hadrons, and more recently of heavy 
flavour hadrons \cite{Hosaka_Prog_Part_Nucl_Phys}
has been a topic of intense research due to the relevance 
to high energy nuclear collision experiments. 
Due to the isospin asymmetry in the heavy colliding nuclei
in the high energy heavy ion collision experiments,
it is important to study the isospin asymmetry 
effects on the properties of the hadrons resulting from
these collisions.
The magnitudes of the magnetic fields created in 
the non-central ultra-relativistic heavy ion collision 
experiments have been estimated to be huge 
($eB \sim 2 m_\pi^2$ at RHIC, BNL, and 
$eB \sim 15 m_\pi^2$ at LHC, CERN)
\cite{HIC_mag_1,HIC_mag_2,HIC_mag_3,HIC_mag_4}. 
This has triggered  extensive studies on the 
effects of strong magnetic fields on the hadron properties
which should show up in the experimental observables
of the heavy ion collision experiments.
The effects of magnetic fields have been studied on the in-medium 
properties of the heavy flavour mesons
\cite{Gubler_D_mag_QSR,machado_1,B_mag_QSR,dmeson_mag,bmeson_mag,charmonium_mag,upsilon_mag,charmonium_mag_QSR,charmonium_mag_lee,charmonium_mag,Suzuki_Lee_2017,Alford_Strickland_2013}.
However, the time evolution of the magnetic field created in
non-central ultra relativistic heavy ion collision experiments
\cite{time_evolution_B_HIC_Tuchin_1,time_evolution_B_HIC_Tuchin_2,time_evolution_B_HIC_Tuchin_3,time_evolution_B_HIC_Tuchin_4,time_evolution_B_HIC_Ajit},
which depends on the electrical
conductivity of the medium and needs solutions of the magnetohydrodynamic
equations, is still an open question. 

The heavy flavour mesons have been studied extensively
in the literature, using the QCD sum rule approach
\cite{kimlee,klingl,amarvjpsi_qsr,jpsi_etac_mag,moritalee_1,moritalee_2,moritalee_3,moritalee_4,open_heavy_flavour_qsr_1,open_heavy_flavour_qsr_2,open_heavy_flavour_qsr_3,open_heavy_flavour_qsr_4,Wang_heavy_mesons_1,Wang_heavy_mesons_2,arvind_heavy_mesons_QSR_1,arvind_heavy_mesons_QSR_2,arvind_heavy_mesons_QSR_3}, 
potential models
\cite{eichten_1,eichten_2,satz_1,satz_2,satz_3,satz_4,satz_5,repko,
Ebert,Bonati_pot_model,Yoshida_Suzuki_heavy_flavour_meson_strong_B}
the coupled channel approach 
\cite{ltolos,ljhs,mizutani_1,mizutani_2,HL,tolos_heavy_mesons_1,tolos_heavy_mesons_2}, the quark meson coupling (QMC) model 
\cite{open_heavy_flavour_qmc_1,open_heavy_flavour_qmc_2,open_heavy_flavour_qmc_3,qmc_1,qmc_2,qmc_3,qmc_4,krein_jpsi,krein_17},
heavy quark symmetry and interaction
of these mesons with nucleons via pion exchange \cite{Yasui_Sudoh_pion},
heavy meson effective theory
\cite{Yasui_Sudoh_heavy_meson_Eff_th}, studying the heavy flavour meson as
an impurity in nuclear matter \cite{Yasui_Sudoh_heavy_particle_impurity}.
The mass modifications of the charmonium states 
have been studied from the medium change of the
scalar gluon condensate calculated in a linear density
approximation in Ref.\cite{leeko}, 
using leading order QCD formula \cite{pes1,pes2,voloshin}. 
Within a chiral effective model \cite{Schechter,paper3,kristof1},
generalized to include the interactions of the charm and bottom
flavoured hadrons,
the in-medium heavy quarkonium (charmonium and bottomonium) masses are
obtained from the medium changes of a scalar dilaton field,
which mimics the gluon condensates of QCD 
\cite{amarvdmesonTprc,amarvepja,AM_DP_upsilon}.
The mass modifications of  the open heavy flavour
(charm and bottom) mesons within the chiral effective model
have also been studied from their interactions with the baryons
and scalar mesons in the hadronic medium
\cite{amdmeson,amarindamprc,amarvdmesonTprc,amarvepja,DP_AM_Ds,DP_AM_bbar,DP_AM_Bs}.
The chiral effective model, in the original version 
with three flavours of quarks (SU(3) model), 
 has been used extensively in the literature,
for the study of finite nuclei \cite{paper3},
strange hadronic matter \cite{kristof1}, 
light vector mesons \cite{hartree}, 
strange pseudoscalar mesons, e.g. the kaons and antikaons
\cite{kaon_antikaon,isoamss,isoamss1,isoamss2}
in isospin asymmetric hadronic matter,
as well as for the study of bulk matter of neutron stars 
\cite{pneutronstar}.
Using the medium changes of the light iuuark condensates
and gluon condensates calculated within the
chiral SU(3) model,
the light vector mesons ($\omega$, $\rho$ and $\phi$) 
in (magnetized) hadronic matter have been 
studied within the framework of QCD sum rule approach 
 \cite{am_vecmeson_qsr,vecqsr_mag}.
The kaons and antikaons have been recently studied
in the presence of strong magnetic fields
using this model \cite{kmeson_mag}.
The model has been used to study the partial decay widths
of the heavy quarkonium states to the open heavy flavour mesons,
in the hadronic medium
\cite{amarvepja} using a light quark creation model \cite{friman},
namely the $^3P_0$ model \cite{3p0_1,3p0_2,3p0_3,3p0_4} as well as 
using a field theoretical model for composite hadrons 
\cite{amspmwg,amspm_upsilon}.
Recently, the effects of magnetic field on the charmonium 
partial decay widths to $D\bar D$ mesons have been
studied using the $^3P_0$ model \cite{charmdecay_mag}
 as well as using the field 
theoretic model of composite hadrons \cite{charmdw_mag}.
In the present work, the effects of magnetic field
on the partial decay width of $\Upsilon (4S)$ to $B \bar B$,
which arise due to their mass modifications,
have been investigated within field theoretical 
model of composite hadrons. 
The masses of the open bottom
($B$ and $\bar B$) mesons as well as the upsilon state
are studied in the magnetized nuclear matter using 
an effective chiral model. Within the model,
the medium modifications of the $B$ and $\bar B$ mesons 
arise due to their interactions with the nucleons and 
the scalar mesons. On the other hand, the mass modification 
of the $\Upsilon$ state is calculated from the 
medium change of a scalar dilaton field, which mimics 
the gluon condensates of QCD. 
For the charged $B^{\pm}$ mesons, there are additional
Landau energy level contributions, which are  not present
for the neutral open bottom ($B^0$ and $\bar {B^0}$) mesons. 
In the presence of the magnetic field, the effects of the pseudoscalar 
meson-vector meson (PV) mixings
\cite{charmonium_mag_lee,charmdw_mag,open_charm_mag_AM_SPM,strange_AM_SPM,Quarkonia_B_Iwasaki_Oka_Suzuki} 
on the bottomonium and open bottom mesons
are also taken into account in the present study. 
These are observed to lead to more significant 
modifications to the masses of the open bottom mesons
(due to $B-B^*$ and $\bar B - \bar {B^*}$ mixings),
as compared to the mass modification of 
the longitudinal component of $\Upsilon (4S)$, 
due to mixing with $\eta_b(4S)$,
for the values of magnetic field considered in the present work.
The dominant contributions from the PV mixing to the masses
of the $B$ and $\bar B$ mesons (along with the Landau level 
contributions to the charged $B^{\pm}$ mesons)
as well as for $\Upsilon(4S)$ meson (due to $\Upsilon(4S)-\eta_b(4S)$
mixing) are observed to lead to significant modifications to the partial 
decay widths of $\Upsilon (4S)$ to the charged and neutral
$B\bar B$ pairs.

The outline of the paper is as follows : In section 2,
we discuss briefly the computation of the in-medium masses of the 
bottomonium state ($\Upsilon(4S)$) and the open bottom 
($B$ and $\bar B$) mesons in magnetized nuclear matter,
using a chiral effective model. The Landau level contributions
(for the charged $B^\pm$ mesons) as well as effects of the
PV mixings on the masses of the $B$, $\bar B$ and 
$\Upsilon (4S)$ are studied in the presence of the external 
magnetic field.
Section 3 gives a brief description of 
the field theoretic model of composite hadrons
with quark (and antiquark) constituents used to
calculate the partial decay widths of $\Upsilon$ to 
$B \bar B$. The decay width is
calculated within the model using the explicit constructions
for the bottomonium state  $\Upsilon (4S)$ as well as the
$B$ and $\bar B$ mesons in terms of the constituent 
quark and antiquark operators and the matrix element of the
quark antiquark pair creation term of the free Dirac Hamiltonian.
The in-medium decay width is calculated from the medium 
modifications of the masses of the decaying bottomonium 
state, as well as the produced $B$ and $\bar B$ mesons
in the magnetized  asymmetric nuclear matter.
In section 4, we discuss the results obtained 
for these in-medium decay widths
in (asymmetric) nuclear matter in presence of strong magnetic fields.
In section 5, we summarize the findings of the present study.

\section{In-medium masses of $\Upsilon (4S)$ and open bottom mesons}

The masses of the $B$ and $\bar B$ mesons in magnetized 
nuclear matter have been studied using a chiral effective model
in Ref. \cite{bmeson_mag}.
The model is a generalization of a chiral SU(3) model,
to include interactions of the open heavy flavour mesons 
with the light hadrons. 
The model incorporates the broken scale invariance of QCD
through a scalar dilaton field, $\chi$, which mimics the gluon
condensate of QCD.
The Lagrangian density of the chiral effective model, in the presence
of a magnetic field, is given as \cite{paper3}
\bea
{\cal L} = {\cal L}_{kin} + \sum_ W {\cal L}_{BW}
          +  {\cal L}_{vec} + {\cal L}_0
+ {\cal L}_{scalebreak}+ {\cal L}_{SB}+{\cal L}_{mag}^{B\gamma},
\label{genlag} \eea
where, $ {\cal L}_{kin} $ corresponds to the kinetic energy terms
of the baryons and the mesons,
${\cal L}_{BW}$ contains the interactions of the baryons
with the meson, $W$ (scalar, pseudoscalar, vector, axialvector meson),
$ {\cal L}_{vec} $ describes the dynamical mass
generation of the vector mesons via couplings to the scalar fields
and contains additionally quartic self-interactions of the vector
fields, ${\cal L}_0 $ contains the meson-meson interaction terms,
${\cal L}_{scalebreak}$ is a scale invariance breaking logarithmic
potential given in terms of a scalar dilaton field, $\chi$ and
$ {\cal L}_{SB} $ describes the explicit chiral symmetry
breaking. The term ${\cal L}_{mag}^{B\gamma}$, describes the interacion
of the baryons with the electromagnetic field, which includes
a tensorial interaction
$\sim \bar {\psi_i} \sigma^{\mu \nu} F_{\mu \nu} \psi_i$,
whose coefficients account for the anomalous magnetic moments
of the baryons \cite{dmeson_mag,bmeson_mag,charmonium_mag}.


Using the mean field approximation, i.e., writing the meson fields
as classical fields, the masses of the $B$ and $\bar B$
mesons are calculated by solving the dispersion relations obtained
from the interaction Lagrangian of these mesons with the 
nucleons and the scalar mesons. The values of the non-strange
($\sigma$) and strange ($\zeta$) scalar-isoscalar meson, 
the scalar-isovector ($\delta$) meson fields, and, the dilaton
field, $\chi$ are obtained by solving their coupled equations
of motion. These are calculated for given values of the 
magnetic field, the baryon density, 
$\rho_B$, and, the isospin asymmetry parameter $\eta$, defined as
$\eta=(\rho_n-\rho_p)/(2\rho_B)$, where, $\rho_p$ and $\rho_n$ are the
number densities of proton and neutron.

The dispersion relations for the $B$ and $\bar B$ mesons are
obtained from the Fourier transformations of the equations of
motion of these mesons. These are given as
\begin{equation}
-\omega^2+ {\vec k}^2 + m_{B(\bar B)}^2
 -\Pi_{B(\bar B)}(\omega, |\vec k|)=0,
\label{dispbbar}
\end{equation}
where $\Pi_{B(\bar B)}$ denotes the self energy
of the $B$ ($\bar B$) meson in the medium,
which are given in terms of the scalar fields,
the number and scalar densities of the nucleons
\cite{DP_AM_bbar,bmeson_mag}.
In the presence of a magnetic field, 
the number and scalar densities of the proton have contributions
from the Landau energy levels \cite{dmeson_mag,bmeson_mag}.
The charged $B^{\pm}$ mesons have contributions from the 
Landau energy levels in the presence of an external magnetic field
\cite{Chernodub,Taya}.
Retaining only lowest Landau level (LLL) contribution 
the mass of the $B^{\pm}$ meson is given as 
\begin{equation}
m^{eff}_{B^\pm}=\sqrt {{m^*_{B^\pm}}^2 +|eB|},
\label{mbpm_landau}
\end{equation}
whereas for the neutral ($B^0$ and $\bar {B^0}$) mesons,
the effective masses are given as
\begin{equation}
m^{eff}_{B^0 (\bar {B^0})}=m^*_{B^0 (\bar {B^0})}.
\label{mb0b0bar}
\end{equation}
It might be noted here that equation (\ref{mbpm_landau})
refers to the mass of a spin zero particle
in a magnetic field due to the lowest Landau level
contribution, ignoring its internal structure
\cite{Chernodub}.
In equations (\ref{mbpm_landau}) and (\ref{mb0b0bar}), 
$m^*_{B^\pm,B^0,\bar {B^0}}$ are the masses calculated 
using the chiral effective model, as the solutions for 
$\omega$ at $|\vec k|=0$, of the dispersion relations 
for these mesons given by equation (\ref{dispbbar}).

The mass shift of the heavy quarkonium state
is proportional to the change in the gluon condensate
in the medium. 
This is the leading order result of a study of the
heavy quarkonium state in a gluon field, assuming  
the distance between the heavy quark and antiquark
(bound by a Coulomb potential)
to be small as compared to the scale of gluonic 
fluctuations \cite{pes1,pes2,voloshin}.
The dilaton field $\chi$ of 
the scale breaking term ${\cal L}_{scalebreak}$ in the chiral
effective model is related to the scalar gluon condensate
of QCD and this relation is obtained by equating the trace
of the energy momentum tensor in the chiral effective model
and in QCD. The mass shift of the bottomonium state 
in the magnetized nuclear matter is hence computed
from the medium change of the dilaton field from vacuum value,
calculated within the chiral effective model,
and is given as \cite{AM_DP_upsilon}
\begin{equation}
\Delta m= \frac{4}{81} (1 - d) \int d |{\bf k}|^{2}
\langle \vert \frac{\partial \psi (\bf k)}{\partial {\bf k}}
\vert^{2} \rangle
\frac{|{\bf k}|}{|{\bf k}|^{2} / m_{b} + \epsilon}
 \left( \chi^{4} - {\chi_0}^{4}\right),
\label{masspsi}
\end{equation}
where
\begin{equation}
\langle \vert \frac{\partial \psi (\bf k)}{\partial {\bf k}}
\vert^{2} \rangle
=\frac {1}{4\pi}\int
\vert \frac{\partial \psi (\bf k)}{\partial {\bf k}} \vert^{2}
d\Omega.
\end{equation}
In equation (\ref{masspsi}), $d$ is a parameter introduced
in the scale breaking term in the Lagrangian,
 $\chi$ and $\chi_0$
are the values of the dilaton field in the magnetized medium 
and in vacuum respectively.
The wave functions of the bottomonium states,
$\psi(\bf k)$ are assumed to be harmonic oscillator
wave functions, $m_b$ is the mass of bottom quark, $\epsilon=2m_b-m$
is the binding energy of the bottomonium state of mass, $m$.

\subsection{Pseudoscalar meson-Vector meson (PV) mixing}

In the presence of a magnetic field, there is mixing between
the pseudoscalar meson and vector mesons, which modifies
the masses of these mesons 
\cite{charmonium_mag_lee,Alford_Strickland_2013,charmdw_mag,open_charm_mag_AM_SPM,strange_AM_SPM,Quarkonia_B_Iwasaki_Oka_Suzuki}.
The PV mixing leads to a drop (rise) in the mass of the
pseudoscalar (longitudinal component of the vector meson).
The mass modifications have been studied 
using an effective Lagrangian density of the form 
${\cal L}_{PV\gamma}\sim {\tilde F_{\mu \nu}}(\partial^\mu P)V^\nu$
\cite{charmonium_mag_lee,Quarkonia_B_Iwasaki_Oka_Suzuki} 
for the heavy quarkonia \cite{charmonium_mag_lee,charmdw_mag},
the open charm mesons \cite{open_charm_mag_AM_SPM}
and strange ($K$ and $\bar K$) mesons
\cite{strange_AM_SPM}.

In the present work, we estimate the modifications to the masses
of the pseudoscalar and vector mesons ($Q_1 {\bar {Q_2}}$ bound states)
due to mixing of these states in the presence of a magnetic field,  
using the Hamiltonian 
\cite{Alford_Strickland_2013,Quarkonia_B_Iwasaki_Oka_Suzuki}.
\begin{equation}
H_{\rm {spin-mixing}}=-{\sum _{i=1}^2} 
{\mbox{\boldmath $\mu$}}_i
\cdot {\bf B},
\label{H_spin_mixing}
\end{equation}
which decribes the interaction of the magnetic 
moments of the quark (antiquark) with the external magnetic field.
In the above, 
${\mbox{\boldmath $\mu$}}_i
=g|e|{q_i} {\bf {S_i}}/(2m_i)$ 
is the magnetic moment of the $i$-th particle, $g$ is the Lande g-factor
(taken to be $2(-2)$ for the quark(antiquark)), $q_i$, $\bf {S_i}$,
$m_i$ are the electric charge (in units of the magnitude of the
electronic charge, $|e|$), spin and mass of the $i$-th particle
\cite{charmonium_mag_lee,Quarkonia_B_Iwasaki_Oka_Suzuki}.
This interaction leads to a drop (increase) of the mass of the 
pseudoscalar (longitudinal component of the vector meson) given as
\cite{Alford_Strickland_2013}
\begin{equation}
{\Delta M}^{PV}= \frac{\Delta E}{2} \Big ( (1+\Delta ^2)^{1/2}-1\Big),
\label{delm_PV}
\end{equation}
where $\Delta=2g|eB|((q_1/m_1)-(q_2/m_2))/\Delta E$,
$\Delta E=m_V-m_P$ is the difference in the masses 
of the pseudoscalar and vector mesons.
As we shall see, the masses of the open bottom ($B$ and $\bar B$) mesons 
are observed to have dominant contributions from the PV 
($B-B^*$ and $\bar B-\bar {B^*}$) mixings,
which, in turn, affect appreciably the partial decay widths $\Upsilon(4S)
\rightarrow B\bar B$ in presence of an external magnetic field.

\section{$\Upsilon (4S) \rightarrow B\bar B$ in a field theoretical model 
for composite hadrons}

We use a field theoretical model for composite hadrons
with quark (and antiquark) constituents \cite{spm781,spm782,spmdiffscat} 
to investigate the
effects of strong magnetic fields on the decay widths
of bottomonium states to $B\bar B$. The model has been used
for calculating the in-medium partial decay widths of the charmonium
(bottomonium) states to $D\bar D$ ($B\bar B$) 
in hot strange hadronic matter, as well as, for
studying the effects of magnetic fields on the
decay widths of charmonium states decaying to $D\bar D$.
In the model, the decay width is calculated 
using explicit constructions for the
heavy quarkonium as well as the open heavy flavour (charm, bottom) 
mesons, and the quark antiquark pair creation term of
the free Dirac Hamiltonian, written in terms of the
constituent quark field operators. The decay amplitude 
is multiplied with a strength parameter for the light quark pair
creation, which is fitted to the vacuum partial decay 
widths of the lowest quarkonium state ($\psi (3770)$ for 
charmonium state and $\Upsilon (4S)$ for bottomonium state) which 
can decay to $D\bar D$ or $B\bar B$ in vacuum.
The present paper investigates 
the partial decay width $\Upsilon (4S)$ to $B\bar B$
in magnetized isospin asymmetric nuclear matter.
The modifications to the decay width are computed from
the mass modifications of the $\Upsilon (4S)$ as well 
as $B$ and $\bar B$ mesons in the magnetized matter.
These mass shifts are calculated within the chiral effective model 
using equations (\ref{masspsi}) and (\ref{dispbbar})
\cite{bmeson_mag,upsilon_mag}. For the charged $B^{\pm}$ mesons
there are (lowest) Landau level contribution, which modifies
their masses as given by equation (\ref{mbpm_landau}).  
The PV mixing effects lead to a drop (rise) in the mass
of the pseudoscalar (longitudinal component of the vector meson)
by ${\Delta M}^{PV}$ given by equation (\ref{delm_PV}).
As we shall see later, the mass modifications of the $B$, $\bar B$ mesons
and $\Upsilon(4S)$ due to the PV mixing are quite significant, 
leading to appreciable modification in the decay widths 
of $\Upsilon (4S)\rightarrow B\bar B$
in the magnetized medium.

For $\Upsilon (4S)$ decaying at rest to $B ({\bf p}) \bar B (-{\bf p})$,
the decay width has already been calculated using the field theoretical
model for composite hadrons with constituent quarks and antiquarks,
in Ref.\cite{amspm_upsilon}. 	
For the sake of completeness, we briefly describe the computation
of the decay width of $\Upsilon (4S) \rightarrow B ({\bf p})
\bar B (-{\bf p})$ in this section. The results obtained
for these decay widths in magnetized nuclear matter will be described
in the next section.
For the decay of $\Upsilon (4S)$ at rest to $B ({\bf p}) \bar B (-{\bf p})$,
the magnitude of $|{\bf p}|$ is given as
\begin{equation}
|{\bf p}|=\Bigg (\frac{{m_\Upsilon}^2}{4}-\frac {{m_B}^2+{m_{\bar B}}^2}{2}
+\frac {({m_B}^2-{m_{\bar B}}^2)^2}{4 {m_\Upsilon}^2}\Bigg)^{1/2}.
\label{pb}
\end{equation}
In the above,  $m_\Upsilon$, $m_B$ and $m_{\bar B}$ are the in-medium 
masses of the bottomonium state ($\Upsilon (4S)$ in the present investigation), 
$B$ and $\bar B$ mesons respectively.

The field operator for a constituent quark for a hadron at rest,
(as the case for the bottomonium state decaying at rest)
at time, t=0, is written as
\begin {equation}
\psi ({\bf x})
=(2\pi)^{-{3}/{2}}{\int \Big [
U({\bf k})
 q_r ({\bf k})u_r
e^{i{\bf k} \cdot{\bf x}}
+ 
V({\bf k}) 
\tilde q_s ({\bf k})v_s
e^{-i{\bf k} \cdot{\bf x}}\Big ]} d{\bf k},
\label{qx}
\end{equation}
where, the operator $q_{r}({\bf k})$ annihilates a quark with spin $r$ 
and momentum ${\bf k}$, whereas, $\tilde q _{s}({\bf k})$
creates an antiquark with spin $s$ and momentum ${\bf k}$,
and these operators satisfy the usual anticommutation relations
\begin{equation}
\{q_{r}({\bf k}),q_{s}({\bf k}')^\dagger\}=
\{\tilde q_{r}({\bf k}),\tilde q_{s}({\bf k}')^\dagger\}=
\delta _{rs} \delta ({\bf k}-{\bf k}').
\end{equation}
In equation (\ref{qx}), 
$U({\bf k})$ and $V({\bf k})$ are given as
\cite{spm781},
\begin{equation}
U({\bfs k})=\left (\begin{array}{c} f(|{{\bf k}}|)\\
{\bfm\sigma}\cdot {\bf k} g(|{{\bf k}}|)\\
\end{array} \right ),\;\;\;\;\;
V({\bfs k})=\left (\begin{array}{c} 
{\bfm\sigma}\cdot {\bf k} g(|{{\bf k}}|)\\
f(|{{\bf k}}|)\\
\end{array} \right ),
\label{ukvk}
\end{equation}
and, $u_r$ and $v_s$ are the two component
spinors for the quark and antiquark.
The functions $f(|{\bf k}|)$ and $g(|{\bf k}|)$ satisfy the constraint
$f^2+g^2 {\bf k}^2=1$,
as obtained from the equal time anticommutation relation 
for the four-component Dirac field operators. 
These functions, for the case of free Dirac field
of mass $M$, are given as,
\begin{equation}
f(|{\bf k}|)=\left ( \frac{k_0 +M}{2 k_0}\right )^{1/2},\;\;\;\; 
g(|{\bf k}|)=\left ( \frac{1}{2 k_0 (k_0+M)}\right )^{1/2},
\label{fkgk}
\end{equation}
where $k_0=(|{\bf k}|^2+M^2)^{1/2}$. In the above, $M$ is the constituent
quark mass, which has been assumed to be momentum independent 
\cite{amspmwg,amspm_upsilon} in the present work.
We also take the low momentum expansions for the
the functions $f(|{\bf k}|) \sim 1-({g(|{\bf k}|)}^2{\bf k}^2/2)$ 
and $g(|{\bf k}|)\sim 1/2M$ \cite{amspmwg,amspm_upsilon}. 

For a hadron in motion (as for the case for
the outgoing $B$ and $\bar B$ mesons),
the field operators for quark annihilation and antiquark creation,
for t=0, are obtained by Lorentz boosting the field operator of the 
hadron at rest. The field operators for the constituent quark
and antiquark of a hadron with four momentum p, 
are given as \cite{spmdiffscat}  
\begin{equation}
Q^{(p)}({\bf x},0)=(2\pi)^{-{3}/{2}} {\int {d\bfs k S(L(p)) U({\bf k}) 
Q_r ({\bf k}+\lambda {\bf p})u_r\exp(i({\bf k}+\lambda {\bf p}) 
\cdot{\bf x})}}
\label{qxp}
\end{equation}
and,
\begin{equation}
\tilde Q^{(p)}({\bf x},0)=(2\pi)^{-{3}/{2}} 
{\int {d\bfs k S(L(p)) V(-{\bf k}) 
\tilde Q_s (-{\bf k}+\lambda {\bf p})v_s
\exp(-i(-{\bf k}+\lambda {\bf p}) \cdot{\bf x})}}.
\label{tldqxp}
\end{equation}
In the above, $\lambda$ is the fraction of the energy of the hadron 
at rest, carried by the constituent quark (antiquark), and, 
the Lorentz boosting factor $S(L(p))$
is given as 
\begin{equation}
S(L(p))=\left ( \frac{p^0+m_h}{2m_h}\right )^{1/2}
+\frac {\bfm\alpha \cdot {\bf p}}{(2m_h (p^0+m_h))^{1/2}},
\label{slp}
\end{equation}
where, ${\bf \alpha}^{i}
=\left (\begin{array}{c} 0 \;\;\; \sigma^i\\
\sigma^i\;\; 0 \end{array}\right )$
 are the Dirac matrices,
and, $m_h$ is the mass of the hadron with momentum ${\bf p}$,
which is the $B(\bar B)$ meson here.

The explicit construct for the state for the bottomonium 
state $\Upsilon (4S)$ with spin projection $m$ at rest,
assuming harmonic oscillator wave functions for the 
bottomonium states, is given as
\begin{equation}
|\Upsilon _m(\vec 0)\rangle = \int {d {\bf k}_1 
b_r^i ({\bf k}_1)^\dagger u_r^\dagger 
u_{\Upsilon(4S)}({\bf k}_1) \sigma_m \tilde {b_s}^i v_s 
(-{\bf k}_1)|vac\rangle},
\label{upsilon}
\end{equation}
where, 
${b_r^i}^\dagger({\tilde b}_s^i)$ creates a bottom 
quark (antiquark) of spin r(s) and color $i$, 
$S_m\equiv \frac{1}{2}\sigma_m$ gives the spin projection
of the bottomonium state,
and, the harmonic oscillator wave function of $\Upsilon(4S)$ 
is given as
\begin{eqnarray}
u_{\Upsilon(4S)}(\bfs k_1)
&=&-\frac{1}{\sqrt{6}}
{\frac{ \sqrt {35}}{4}}
\Bigg({\frac{R_{\Upsilon(4S)}^2}{\pi}}\Bigg)^{3/4}
\left(1-2 R_{\Upsilon(4S)}^2\bfs k_1^2
+\frac{4}{5}R_{\Upsilon(4S)}^4\bfs k_1^4
-\frac{8}{105}R_{\Upsilon(4S)}^6\bfs k_1^6
\right)\nonumber \\
&\times &
\exp\left[-\frac{1}{2}R_{\Upsilon(4S)}^2\bfs k_1^2\right].
\label{u4s}
\end{eqnarray}
In the above, $R_{\Upsilon (4S)}$ is the 
strength parameter of the harmonic oscillator wave function,
$u_{\Upsilon (4S)}$.
The factor $\frac {1}{\sqrt 6}$ in equation (\ref{u4s}) 
refers to normalization
factor arising from degeneracy factors due to color (3) and 
spin (2)  of the quarks and antiquarks.
The $B^0$ and $\bar {B^0}$ states, with finite momenta, 
are explicitly given as
\begin{equation}
|B^0 ({\bf p}')\rangle = { \int {d_r^{i_2}({\bf k}_3
+\lambda_1 {\bf p}') ^\dagger u_r^\dagger
 u_{B}({\bf k}_3)\tilde {b_s}^{i_2} 
(-{\bf k}_3 +\lambda_2 {\bf p}') v_s 
d\bfs k_3}}.
\label{b0}
\end{equation}
and 
\begin{equation}
|\bar {B^0} ({\bf p})\rangle 
={ \int {b_r^{i_1}}({\bf k}_2+\lambda_2 {\bf p}) ^\dagger 
u_r^\dagger
u_{B}({\bf k}_2)\tilde {d_s}^{i_1} (-{\bf k}_2 +\lambda_1 {\bf p}) 
v_s d\bfs k_2},
\label{b0bar}
\end{equation}
where, 
\begin{equation}
u_{B}({\bf k})=\frac{1}{\sqrt{6}}\Bigg 
(\frac {{R_B}^2}{\pi} \Bigg)^{3/4}
\exp\Bigg(-\frac {{R_B}^2 {{\bf k}}^2}{2}\Bigg),
\label{ub}
\end{equation}
is the wave function of $B(\bar B)$ with $R_B$ as the 
strength parameter of the harmonic oscillator wave function.
The charged states $B^-$ ($B^+$) mesons at finite momenta
are obtained by replacing the $d^\dagger (d)$ in the states for $B^0$ 
($\bar {B^0}$) by $u^\dagger (u)$.
In the above, $\lambda_1$ and $\lambda_2$ correspond to the 
fractions of the energy of the hadron carried by
the constituent quark (antiquark). These are determined 
by assuming that the binding energy of the hadron 
shared by the quark (antquark) is inversely 
proportional to the quark (antiquark) mass
\cite{spm782,amspmwg,amspm_upsilon}. 

The matrix element of the quark-antiquark pair creation
part of the Dirac Hamiltonian density,
between the initial and the final states, $M_{fi}$
is evaluated to calculate the partial decay width
for the reaction
$\Upsilon (4S) \rightarrow {\bar B}^0 ({\bf p})+{B^0}(-{\bf p})$.
The expression obtained for the partial decay width 
of the bottomonium state, $\Upsilon (4S)$
decaying at rest to $B^0{\bar {B^0}}$ pair, after averaging
$|M_{fi}|^2$ over spin, is given as \cite{amspmwg}, 
\begin{eqnarray}
\Gamma(\Upsilon (4S) \rightarrow B^0 {\bar {B^0}})
&=& \gamma_\Upsilon ^2 \frac{1}{2\pi} 
\int \delta(m_{\Upsilon (4S)}-p^0_{B^0}-p^0_{\bar {B^0}})
{|M_{fi}|^2}_{av}
\cdot 4\pi |{\bfs p}_{B^0}|^2 d|{\bfs p}_{\bar {B^0}}| 
\nonumber\\
&=& \gamma_\Upsilon ^2\frac{8\pi^2}{3}|{\bf p}|^3
\frac {p^0_{B^0} p^0_{\bar {B^0}}}{m_{\Upsilon (4S)}}
A^{\Upsilon (4S)}(|{\bf p}|)^2
\label{gammaupslnb0b0b}
\end{eqnarray}
In the above, $p^0_{B^0}=\big(m_{B^0}^2+|{\bf p}|^2\big)^{{1}/{2}}$, 
$p^0_{\bar {B^0}}=\big(m_{\bar {B^0}}^2+|{\bf p}|^2\big)^{{1}/{2}}$, 
and, $|\bfs p|$ is the magnitude of the momentum of the outgoing 
$B^0 (\bar {B^0})$ mesons. The decay of 
$\Upsilon (4S)$ to $B^+ B^-$ proceeds through a 
$u \bar u$ pair creation and the decay width 
(\ref{gammaupslnb0b0b}) is modified to 
\begin{eqnarray}
\Gamma(\Upsilon (4S) \rightarrow B^+ B^-)
=&& \gamma_\Upsilon^2\frac{8\pi^2}{3}\cdot|{\bf p}|^3
\frac {{p^0}_{B^+} {p^0}_{B^-}}{m_{\Upsilon (4S)}}
A^{\Upsilon (4S)}(|{\bf p}|)^2
\label{gammaupslnbpbm}
\end{eqnarray}
In the above, $p^0_{B^\pm}=\big(m_{B^\pm}^2+|{\bf p}|^2\big)^{{1}/{2}}$, 
and, $|\bfs p|$ is the magnitude of the momentum of the outgoing 
$B^\pm$ mesons. The expression for 
$A^{\Upsilon (4S)}(|{\bf p}|)$ is given as
\begin{eqnarray}
A^{\Upsilon (4S)}(|{\bf p}|) &= & 
6c_{\Upsilon (4S)}\exp[(a_{\Upsilon (4S)} 
{b^2_{\Upsilon (4S)}}
-R_B^2\lambda_2^2)|{\bf p}|^2]
\cdot\Bigg(\frac{\pi}{a_{\Upsilon (4S)}}\Bigg)^{{3}/{2}}
\nonumber \\
&\times &
\Bigg[F_0^{\Upsilon (4S)}+\frac{3}{2a_{\Upsilon (4S)}}
\cdot F_1^{\Upsilon (4S)}
+ \frac{15}{4a_{\Upsilon (4S)}^2}\cdot F_2^{\Upsilon (4S)}
\nonumber \\
&+&
\frac{105}{8a_{\Upsilon (4S)}^3} \cdot F_3^{\Upsilon (4S)}
+ \frac{105 \times 9}{16a_{\Upsilon (4S)}^4}
\cdot F_4^{\Upsilon (4S)}
\Bigg].
\label{ap}
\end{eqnarray}
In the above equation, $F_{i}^{\Upsilon (4S)} (|{\bf p}|)$'s, 
$i=0,1,2,3,4$, have been computed in Ref. \cite{amspm_upsilon},
which for sake of completeness, we quote in the following.
\begin{eqnarray}
F^{\Upsilon (4S)}_0
&=& \frac {1}{2} (b_{\Upsilon (4S)}-1) (b_{\Upsilon (4S)}-\lambda_2)
(3 b_{\Upsilon (4S)} +\lambda_2 -4) g^2 |{\bf p}|^2
\nonumber \\
&&\times  
\Bigg (1-2 R_{\Upsilon (4S)}^2 
b_{\Upsilon (4S)}^2 |{\bf p}|^2
+\frac{4}{5} R_{\Upsilon (4S)}^4 
b_{\Upsilon (4S)}^4 {|\bf p|}^4 \nonumber \\
&-&\frac{8}{105} R_{\Upsilon (4S)}^6 
b_{\Upsilon (4S)}^6 {|\bf p|}^6
\Bigg)
\end{eqnarray}
\begin{eqnarray}
F^{\Upsilon (4S)}_1
&=&
\frac {g^2}{6} \Big ( 9 (b_{\Upsilon (4S)}-1)
- 2 (3 b_{\Upsilon (4S)} -\lambda_2 -2) \Big )
\nonumber \\
&+& \frac {g^2 {|\bf p|}^2  R_{\Upsilon (4S)}^2}{3}
\Bigg [ (-5 b_{\Upsilon (4S)} +3) (3b_{\Upsilon (4S)} 
+\lambda_2 -4 )(b_{\Upsilon (4S)}-\lambda_2) 
\nonumber \\
&-& 
9 b_{\Upsilon (4S)} ^2 (b_{\Upsilon (4S)}-1) 
+ 2 b_{\Upsilon (4S)} (3 b_{\Upsilon (4S)} -\lambda_2 -2) 
( 3 b_{\Upsilon (4S)} -2)
\Bigg ]
\nonumber \\
&+&\frac {4 g^2 {|\bf p|}^4 R_{\Upsilon (4S)} ^4 b_{\Upsilon (4S)}^2}{15}
\Bigg [ (7b_{\Upsilon (4S)} -5) (3b_{\Upsilon (4S)} 
+\lambda_2 -4 )(b_{\Upsilon (4S)}-\lambda_2) 
\nonumber \\
&+& \frac {9}{2} (b_{\Upsilon (4S)} -1) b_{\Upsilon (4S)} ^2 
-  b_{\Upsilon (4S)} ( 5 b_{\Upsilon (4S)} - 4)
(3b_{\Upsilon (4S)} -\lambda_2 -2 )\Bigg ]
\nonumber \\
&-& 
\frac {8 g^2 {|\bf p|}^6 R_{\Upsilon (4S)} ^6 b_{\Upsilon (4S)}^4}{105}
\Bigg [ \frac {1}{2}(9b_{\Upsilon (4S)} -7) 
(3b_{\Upsilon (4S)} +\lambda_2 -4 )(b_{\Upsilon (4S)}-\lambda_2) 
\nonumber \\
&+&\frac {3}{2} b_{\Upsilon (4S)} ^2 (b_{\Upsilon (4S)}-1) 
-\frac {1}{3} b_{\Upsilon (4S)} (3 b_{\Upsilon (4S)} 
-\lambda_2 -2) (7b_{\Upsilon (4S)} -6)
\Bigg ]
\end{eqnarray}
\begin{eqnarray}
 F^{\Upsilon (4S)}_2
&=&\frac {1}{3} g^2 R_{\Upsilon (4S)}^2 (-9 b_{\Upsilon (4S)} 
-2 \lambda_2 +5)
\nonumber \\
&+& \frac {4}{5} g^2 R_{\Upsilon (4S)} ^4 |{\bf p}|^2
\Bigg [ b_{\Upsilon (4S)} ^2 (7 b_{\Upsilon (4S)} -5) 
\nonumber \\
&+& \frac {1}{6} (3 b_{\Upsilon (4S)} +\lambda_2 -4)
(b_{\Upsilon (4S)} -\lambda_2) (7 b_{\Upsilon (4S)} -3) 
\nonumber \\
 &-& \frac {2}{15}  b_{\Upsilon (4S)} 
(3 b_{\Upsilon (4S)} -\lambda_2 -2)
(21 b_{\Upsilon (4S)} -10) \Bigg ]
\nonumber \\
&+& \frac {4}{5} g^2 R_{\Upsilon (4S)} ^6 |{\bf p}|^4 b_{\Upsilon (4S)} ^ 2
\Bigg [ -\frac {1}{7} b_{\Upsilon (4S)}^2 (9 b_{\Upsilon (4S)} -7)
\nonumber \\
&-&\frac {4}{15} b_{\Upsilon (4S)} (b_{\Upsilon (4S)} -\lambda_2 ) 
(3 b_{\Upsilon (4S)} +\lambda_2 -4)
\nonumber \\
&& -\frac {1}{3} (b_{\Upsilon (4S)}-1) (b_{\Upsilon (4S)} -\lambda_2) 
(3 b_{\Upsilon (4S)} +\lambda_2 -4) 
\nonumber \\
&+ & \frac {2}{105} b_{\Upsilon (4S)} (3b_{\Upsilon (4S)} -\lambda_2 -2) 
(45 b_{\Upsilon (4S)} -28 ) \Bigg ],  
\end{eqnarray}
\begin{eqnarray}
 F^{\Upsilon (4S)}_3
&=&\frac {2 g^2}{15} R_{\Upsilon (4S)} ^4 (15 b_{\Upsilon (4S)} 
+2 \lambda_2 -5)
\nonumber \\
&+&\frac {4}{5} g^2 R_{\Upsilon (4S)} ^6 |{\bf p}|^2
\Bigg [ -\frac {4}{5} b_{\Upsilon (4S)}^3 
- (b_{\Upsilon (4S)}-1) b_{\Upsilon (4S)} ^2 
\nonumber \\
&-&\frac {2}{21} b_{\Upsilon (4S)} 
(b_{\Upsilon (4S)}-\lambda_2) (3 b_{\Upsilon (4S)} +\lambda_2 -4)
\nonumber \\
&-&\frac {1}{21} (b_{\Upsilon (4S)}-1) (b_{\Upsilon (4S)}-\lambda_2) 
(3 b_{\Upsilon (4S)} +\lambda_2 -4)) 
\nonumber \\
&+& \frac {2}{105} b_{\Upsilon (4S)} (3 b_{\Upsilon (4S)} -\lambda_2 -2)
(27b_{\Upsilon (4S)} -10) \Bigg ] ,
\end{eqnarray}
\begin{eqnarray}
 F^{\Upsilon (4S)}_4
&=&-\frac {4 g^2 R_{\Upsilon (4S)} ^6 }{35\times 9} 
(21b_{\Upsilon (4S)} + 2 \lambda_2 -5).
\label{fiupsln4s}
\end{eqnarray}
\noindent In the above, the parameters $a_{\Upsilon (4S)}$ and 
$b_{\Upsilon (4S)}$
are given as
\begin{equation}
a_{\Upsilon (4S)}=\frac{1}{2}R_{\Upsilon (4S)}^2+R_B^2; \;\;\;\; 
b_{\Upsilon (4S)}=R_B^2\lambda_2/a_{\Upsilon (4S)},
\label{abupsln}
\end{equation}
with $R_{\Upsilon (4S)}$ as the radius of the bottomonium state,
$\Upsilon (4S)$, and,
\begin{equation}
c_{\Upsilon (4S)}=\frac{1}{6\sqrt{6}}
\left(\frac{\sqrt {35}}{4}\right)
\left(\frac{R_{\Upsilon(4S)}^2}{\pi}\right)^{{3}/{4}}
\cdot\left(\frac{R_B^2}{\pi}\right)^{{3}/{2}}.
\end{equation}
The parameter $\gamma_\Upsilon$, has been introduced in the
expressions for the decay widths of $\Upsilon (4S) \rightarrow
B^0 \bar {B^0} (B^+ B^-)$, which is a measure of
the production strength of the $B\bar B$ pair from
the $\Upsilon$-state through light quark antiquark pair 
($d\bar d$ or $u\bar u$) creation. 
In the present investigation of the bottomonium 
decay widths, the parameter, $\gamma_\Upsilon$
is fitted from the vacuum decay width for the
channel $\Upsilon (4S) \rightarrow B\bar B$
($\Upsilon (4S)$ is the lowest $\Upsilon$-state
which decays to $B\bar B$ in vacuum).
The decay widths of the bottomonium state 
depend on the magnitude of $B(\bar B)$ meson momentum, 
$|{\bf p}|$ as a polynomial function
multiplied by an exponential factor, as can be seen from the
expressions given by equations (\ref{gammaupslnb0b0b})
and (\ref{gammaupslnbpbm}). The medium modification
of the decay width is due to the mass changes of the
bottomonium state and the open bottom mesons, which 
are incorporated in  $|{\bf p}|$ given by equation 
(\ref{pb}).

When we include the PV mixing effect (for the $\Upsilon(4S)$ with
the pseudoscalar meson, $\eta_b(4S)$), the mass of the longitudinal
component of $\Upsilon(4S)$ is modified, and, the expression
for the decay width of $\Upsilon(4S)\rightarrow B\bar B$
is given as \cite{charmdw_mag}
\begin{eqnarray}
&&\Gamma^{PV}(\Upsilon (4S) \rightarrow  B({\bf p}) {\bar B} (-{\bf p}))
=\gamma_\Upsilon^2\frac{8\pi^2}{3}
\Bigg [
\Bigg(\frac{2}{3} |{\bf p}|^3
\frac {p^0_B (|{\bf p}|) p^0_{\bar B}(|{\bf p}|)}{m_{\Upsilon (4S)}}
A^{\Upsilon(4S)}(|{\bf p}|)^2 \Bigg)
\nonumber \\
&+&\Bigg(\frac{1}{3} |{\bf p}|^3
\frac {p^0_B(|{\bf p}|) p^0_{\bar B}(|{\bf p}|)}{m_{\Upsilon(4S)}^{PV}}
A^{\Upsilon(4S)}(|{\bf p}|)^2 \Bigg) \Big({|{\bf p}|\rightarrow |{\bf p|}
(m_{\Upsilon(4S)} = m_{\Upsilon(4S)}^{PV})}\Big)
\Bigg].
\label{gammaupsilonbbar_mix}
\end{eqnarray}
In the above, the first term corresponds to the transverse
polarizations for the bottomonium state, $\Upsilon(4S)$,
whose masses remain unaffected by the mixing of the
pseudoscalar and vector bottomonium
states. The second term in (\ref{gammaupsilonbbar_mix})
corresponds to the longitudinal component of the
$\Upsilon (4S)$ state whose mass is modified due to mixing
with the pseudoscalar meson, $\eta_b(4S)$, in the presence of the
magnetic field.

\section{Results and Discussions}

\begin{figure}
\includegraphics[width=17cm,height=17cm]{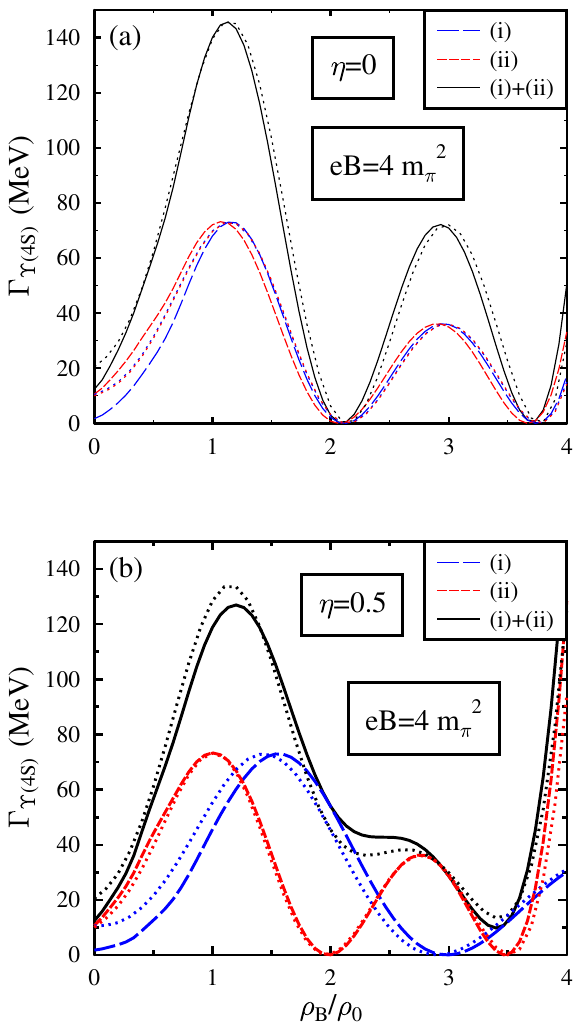}
\vskip -0.5in
\caption{
(Color online)
The decay widths for the bottomonium state $\Upsilon(4S)
\rightarrow B\bar B$ plotted as functions of density
for $\eta=0$ and $\eta=0.5$ and for $eB=4 m_\pi^2$.
The decay widths for the sub-processes (i) 
$\Upsilon (4S) \rightarrow B^+B^-$ and (ii)
$\Upsilon (4S) \rightarrow B^0 \bar {B^0}$,
as well as (iii) the total of these processes
are shown. These decay widths are compared with
the zero magnetic field case, shown as dotted lines.
The mass modifications due to PV mixing effects have not 
been considered.
}
\label{dwFT_amm_eb4}
\end{figure}
\begin{figure}
\includegraphics[width=17cm,height=17cm]{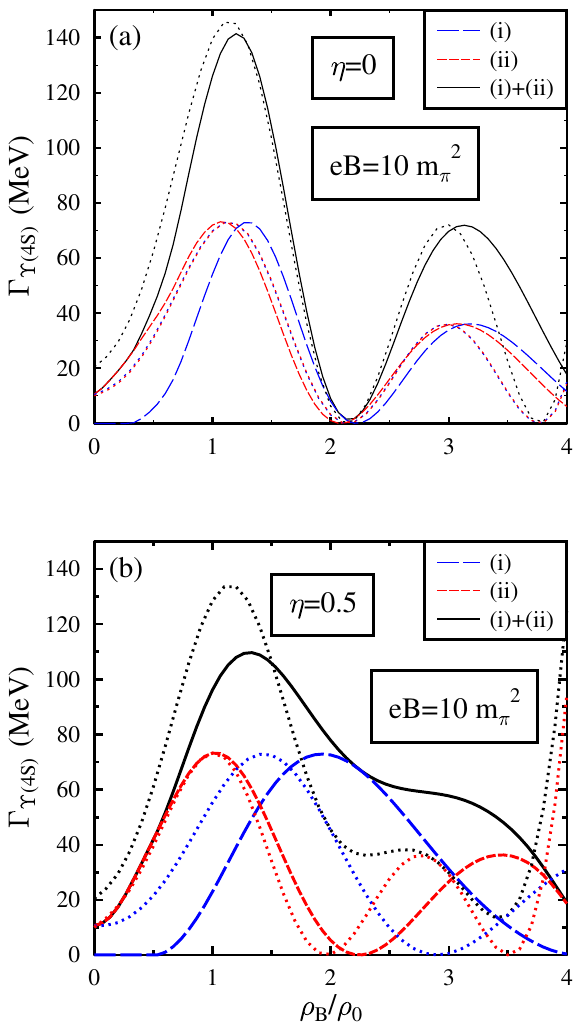}
\vskip -0.5in
\caption{(Color online)
The decay widths for the bottomonium state $\Upsilon(4S)
\rightarrow B\bar B$ plotted as functions of density
for $\eta=0$ and $\eta=0.5$ and for $eB=10 m_\pi^2$.
The decay widths for the sub-processes (i) 
$\Upsilon (4S) \rightarrow B^+B^-$ and (ii)
$\Upsilon (4S) \rightarrow B^0 \bar {B^0}$,
as well as (iii) the total of these processes
are shown. These decay widths are compared with
the zero magnetic field case, shown as dotted lines.
The mass modifications due to PV mixing effects have not 
been considered.
}
\label{dwFT_amm_eb10}
\end{figure}

\begin{figure}
\hskip -0.93in
\includegraphics[width=15cm,height=15cm]{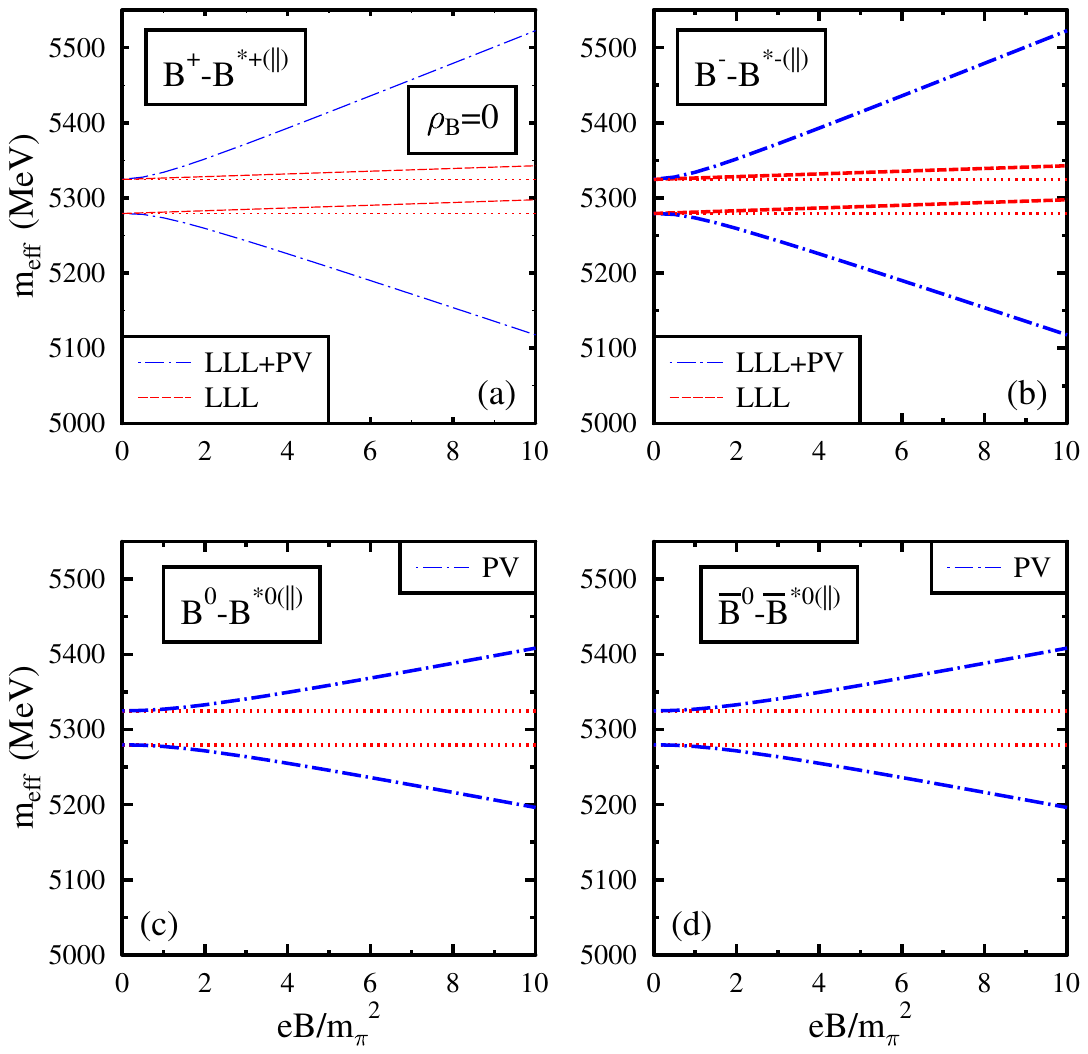}
\vskip -0.5in
\caption{
(Color online)
The masses of the $B^+,B^-,B^0$ and $\bar {B^0}$ 
(and the longitudinal components of their vector meson 
counterparts $B^{*+},B^{*-}, 
B^{*0}$ and ${\bar B}^{*0}$) modified due to the PV mixing effects
(along with LLL contributions for the charged mesons)
are plotted for $\rho_B=0$ as functions of $eB/m_\pi^2$ 
in (a), (b), (c) and (d) respectively. 
The dotted lines illustrate the masses when the LLL (for the charged
mesons) and PV effects are not taken into account. 
}
\label{mbbar_PV_rhbr0}
\end{figure}

\begin{figure}
\hskip -0.93in
\includegraphics[width=15cm,height=15cm]{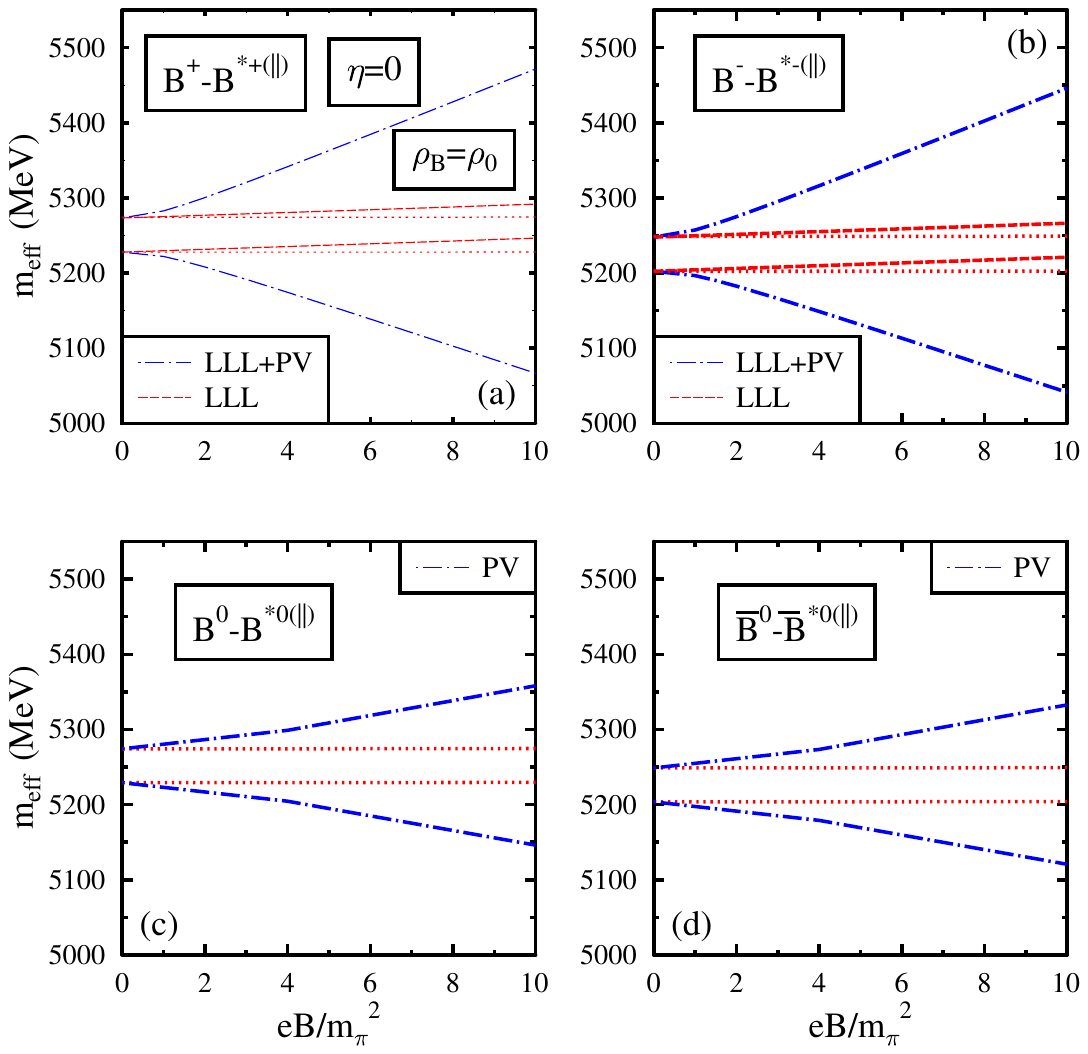}
\vskip -0.5in
\caption{
(Color online)
The masses of the $B^+,B^-,B^0$ and $\bar {B^0}$ 
(and the longitudinal components of their 
vector meson counterparts $B^{*+},B^{*-}, 
B^{*0}$ and ${\bar B}^{*0}$) modified due to the PV mixing effects
(along with LLL contributions for the charged mesons)
are plotted for $\rho_B=\rho_0$ in magnetized symmetric ($\eta$=0) 
nuclear matter
as functions of $eB/m_\pi^2$ in (a), (b), (c) and (d) respectively. 
The dotted lines illustrate the masses as calculated within 
the chiral effective model, which show marginal dependence
on the magnetic field.
}
\label{mbbar_PV_rh0_eta0}
\end{figure}

\begin{figure}
\hskip -0.93in
\includegraphics[width=15cm,height=15cm]{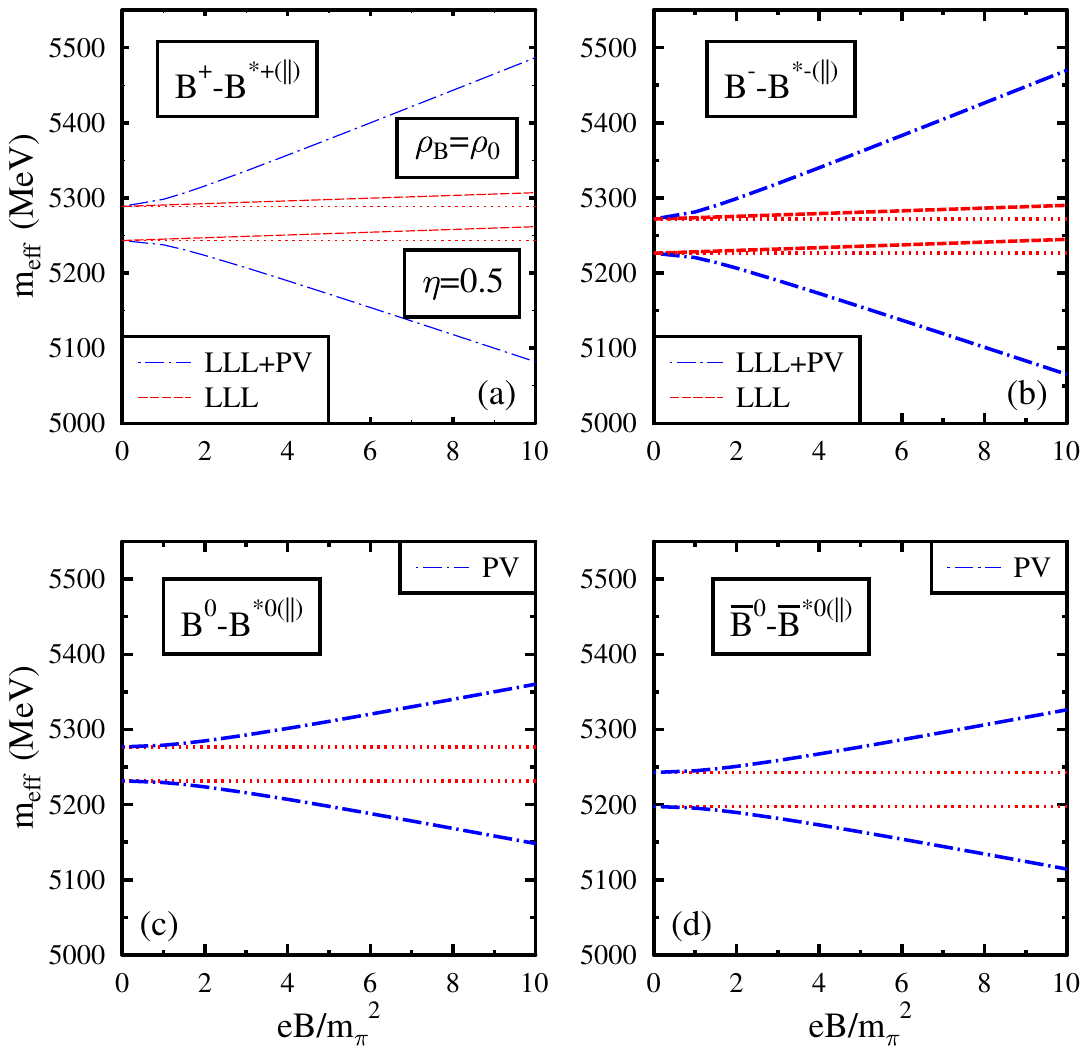}
\vskip -0.5in
\caption{
(Color online)
The masses of the $B^+,B^-,B^0$ and $\bar {B^0}$ 
(and the longitudinal components of their vector meson 
counterparts $B^{*+},B^{*-}, 
B^{*0}$ and ${\bar B}^{*0}$) modified due to the PV mixing effects
(along with LLL contributions for the charged mesons)
are plotted for $\rho_B=\rho_0$ in magnetized asymmetric 
($\eta$=0.5) nuclear matter
as functions of $eB/m_\pi^2$ in (a), (b), (c) and (d) respectively. 
The dotted lines illustrate the masses as calculated within 
the chiral effective model, which show marginal dependence
on the magnetic field.
}
\label{mbbar_PV_rh0_eta5}
\end{figure}

\begin{figure}
\hskip -1.35in
\includegraphics[width=16cm,height=16cm]{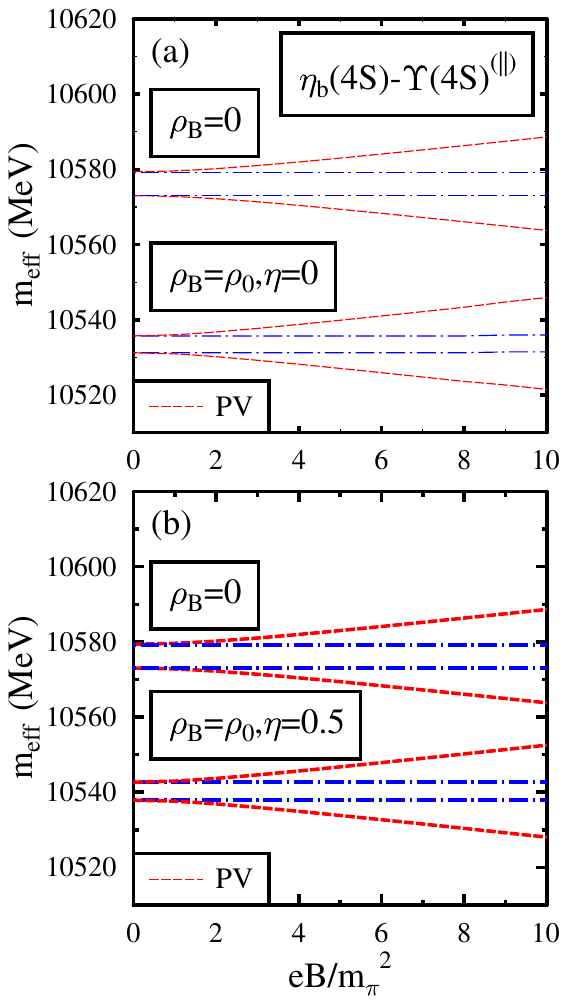}
\vskip -0.5in
\caption{
(Color online)
The masses of the $\Upsilon(4S)$ and $\eta_b(4S)$
with and without the PV mixing effects,
are plotted as functions of $eB/m_\pi^2$. 
These are shown for $\rho_B=\rho_0$ in magnetized symmetric 
($\eta$=0) and in asymmetric ($\eta$=0.5) nuclear matter
in (a) and (b) respectively. These masses are compared
to the effects of PV mixing for the case of $\rho_B=0$.
}
\label{mupsln_etab_4s}
\end{figure}
\begin{figure}
\hskip -0.924in
\includegraphics[width=15cm,height=15cm]{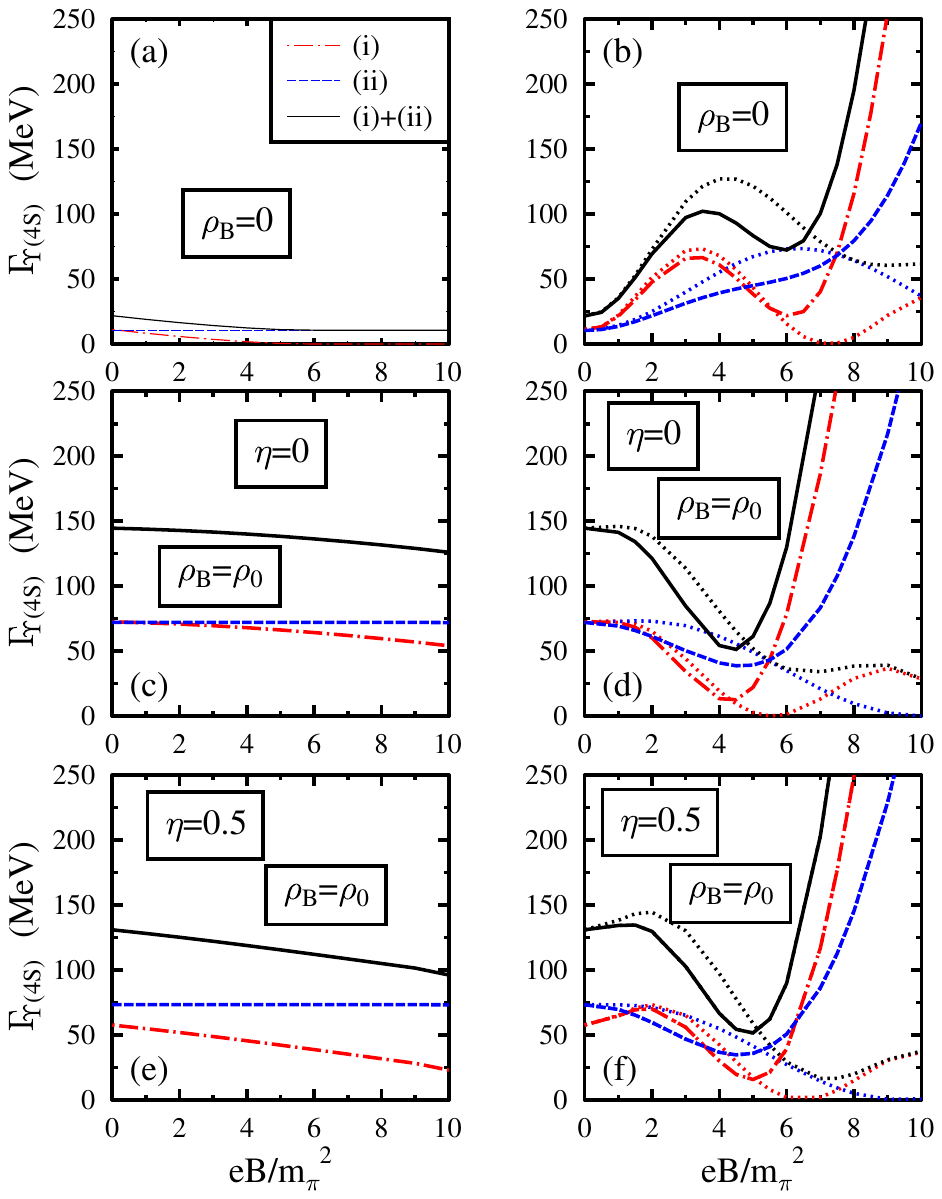}
\vskip -0.5in
\caption{
(Color online)
The decay widths for $\Upsilon (4S)$ to (i) $B^+ B^-$, 
(ii) $B^0 \bar {B^0}$, and the total of these two processes
((i)+(ii)) are plotted as functions of $eB/m_\pi^2$ 
for $\rho_B$=0, as well as, for $\rho_B=\rho_0$ for
symmetric ($\eta$=0) and asymmetric  ($\eta$=0.5) 
nuclear matter in the presence of a magnetic field. 
In (a), (c) and (e), these decay widths
are shown for the case when PV mixing effects are not taken
into account, but the lowest Landau level (LLL)
contributions are taken into account for the charged
$B^{\pm}$ mesons. Subplots (b), (d) and (f) show
the decay widths including the PV mixing effects,
where the dotted lines correspond to the case 
when the mass modification of $\Upsilon(4S)$
from PV mixing is not considered.
\label{dwFT_upsilon_4s_etab4s_rh0}
}
\end{figure}

The in-medium decay widths of $\Upsilon (4S) \rightarrow B\bar B$,
are calculated using a field theoretic model for composite hadrons 
in the present investigation. The model was used to study the 
partial decay widths of charmonium states to $D\bar D$
in (magnetized) hadronic matter \cite{amspmwg,charmdw_mag},
as well as, of bottomonium states to $B\bar B$
in hadronic matter, at zero magnetic field \cite{amspm_upsilon}.
The present work accounts for the effects of
an external magnetic field for the study of the
in-medium decay widths of $\Upsilon (4S)$
to $B\bar B$, which is the lowest bottomonium state
which can decay to $B\bar B$ in vacuum. 
The anomalous magnetic moments
for the nucleons have been taken into account
\cite{broderick1,broderick2,Wei,mao,amm,VD_SS_1,VD_SS_2,aguirre_fermion}
in the present work. 

In the present calculations, we take the parameters as follows. 
The values of the constituent masses (in MeV) of the $u(d)$ and
bottom quarks are taken to be $m_{u(d)}$=330, $m_b$=5360.
The harmonic oscillator strength of $\Upsilon(4S)$ is fitted
from its observed decay width of 0.272 keV, to be
$R_{\Upsilon(4S)}^{-1}=638.6$ MeV \cite{amspm_upsilon}. 
For the $B(\bar B)$ mesons,
this strength is taken as $R_B^{-1}$=875.6 MeV,
assuming $R_B=R_D(m_D/m_B)$, with the harmonic oscillator strength
for $D$ meson obtained as fitted from the partial decay widths 
of $\psi(3770)$ and $\psi(4040)$ to the open charm mesons
\cite{leeko,amspmwg}. The value of the strength parameter,
$\gamma_\Upsilon$ is obtained to be 5.6, as fitted from 
the observed decay widths
of $\Upsilon(4S)\rightarrow B^+B^-$ and
$\Upsilon(4S)\rightarrow B^0{\bar {B^0}}$
as 10.516 MeV and 9.984 MeV respectively \cite{amspm_upsilon}.
The mass modification of $\Upsilon (4S)$ due to its
mixing with $\eta_b(4S)$ in the presence of a magnetic field
is considered in the present study.
The harmonic oscillator strength of $\eta_b(4S)$, 
is calculated to be $R_{\eta_b(4S)}^{-1}=642.627$ MeV 
from a linear interpolation 
along with $\Upsilon(3S)$ and $\Upsilon(4S)$ in their
mass vs $(R)^{-1}$ plot, where the values of
$R^{-1}$ for $\Upsilon(3S)$
and $\Upsilon (4S)$ as fitted from their leptonic decay widths
are 779.75 MeV and 638.60 MeV respectively \cite{amspm_upsilon}. 

The decay widths $\Upsilon(4S)\rightarrow B\bar B$,
in magnetized nuclear matter, along with 
the decay widths for the sub-processes
(i) $\Upsilon (4S) \rightarrow B^+ B^-$ and 
(ii) $\Upsilon (4S) \rightarrow B^0 \bar {B^0}$,
are plotted in figures \ref{dwFT_amm_eb4} and  \ref{dwFT_amm_eb10}
for magnetic fields, $eB=4 m_\pi^2$ and  $eB=10 m_\pi^2$ 
respectively. 
The decay width to the charged $B\bar B$ mesons
accounts for the Landau level contributions to the masses
of $B^\pm$ mesons. However, the decay widths plotted 
in figures \ref{dwFT_amm_eb4} and \ref{dwFT_amm_eb10}
do not consider the effects due to
PV mixing on the masses of the decaying particle,
$\Upsilon (4S)$, as well as the $B$ and $\bar B$ mesons.
The density dependences of these widths
are shown in these figures for the isospin symmetric 
nuclear matter ($\eta$=0) as well as for the case 
of asymmetric nuclear matter (with $\eta$=0.5). 
These are compared with the case of zero magnetic field,
shown as dotted lines.

In the presence of the magnetic field, the proton,
being electrically charged, has contributions 
from the Landau energy levels. The effects of the
magnetic field are also taken into account 
in the present work by considering the anomalous
magnetic moments of the nucleons. In isospin symmetric
nuclear matter ($\rho_p=\rho_n$), the presence 
of magnetic field thus introduces differences
in the mass modifications between the $B^0$ and
$B^+$ within the $B$ doublet as well as between
$B^-$ and $\bar {B^0}$ within the $\bar B$ doublet.
The difference in the masses have additional 
positive mass shifts for the charged $B^\pm$
in the magnetized matter, arising from the lowest
Landau level (LLL). The positive mass shifts
of the $B^\pm$ due to LLL are observed to lead to suppression
of the decay of $\Upsilon (4S)$ to $B^+ B^-$,
as compared to the decay to the neutral $B^0 \bar {B^0}$,
as can be seen from the figures \ref{dwFT_amm_eb4}
as well as \ref{dwFT_amm_eb10}. For $eB=4 m_\pi^2$,
as can be seen from figure \ref{dwFT_amm_eb4},
at zero density, the decay to $B^+ B^-$ is negligible
($\sim$ 1.7 MeV)
and the total decay width for $\Upsilon (4S) \rightarrow
B\bar B$ ($\sim$ 12.4 MeV) is due to the decay mode to the neutral 
$B\bar B$ pair ($\sim$ 10.6 MeV). On the other hand, in vacuum,
the partial decay widths for $\Upsilon (4S) \rightarrow
B^+ B^-$ and $B^0 \bar {B^0}$, are 10.5 MeV and 10 MeV 
respectively, with total decay width of 20.5 MeV.
In isospin symmetric nuclear matter in presence of 
magnetic field of $eB= 4 m_\pi^2$, the in-medium decay widths 
for the channels $\Upsilon (4S) \rightarrow B^+B^-$,
as well as $\Upsilon (4S) \rightarrow B^0 \bar {B^0}$, 
are observed to have an initial increase with density,
followed by a drop leading to vanishing of the decay widths
at around 2.1 $\rho_0$. Above this density, there is again 
observed to be an increase in the decay widths 
followed by decrease with subsequent vanishing of these decay
widths at around a density of 3.8$\rho_0$. These behaviours
were observed for the case of zero magnetic field, and the 
values of the densities where the decay widths vanish,
were observed to be the same. 
In the asymmetric nuclear medium (with $\eta$=0.5), 
the partial decay width for the decay mode 
$\Upsilon (4S) \rightarrow B^+B^-$ is observed to vanish
at a density of around 3$\rho_0$, whereas for the
neutral $B\bar B$ pair, the decay width is zero
at densities of 2 $\rho_0$ as well as 3.5$\rho_0$.

The effects of the magnetic field on the decay widths
of $\Upsilon (4S) \rightarrow B\bar B$ are observed to be much larger,
as might be seen from figure \ref{dwFT_amm_eb10}, which is plotted
for $eB=10 m_\pi^2$. The value
of the density above which the decay of $\Upsilon (4S) \rightarrow
B^+B^-$ becomes possible, is observed to be around 0.4 (0.5) $\rho_0$ 
for symmetric (asymmetric) nuclear matter for $eB=10 m_\pi^2$.
The effects of the isopsin asymmetry on the decay
widths of $\Upsilon (4S) \rightarrow B \bar B$ are observed to be much
larger at higher densities for the higher magnetic field
of $eB=10 m_\pi^2$, as can be seen from figure 
\ref{dwFT_amm_eb10}. The decay of $\Upsilon (4S) \rightarrow
B^+B^-$ is suppressed as compared to the decay to $B^0 \bar {B^0}$
at low densities, which is due to the increase in the masses
of the charged $B^\pm$ mesons arising from the Landau level 
contributions. 

The masses of the open bottom pseudoscalar mesons ($B^+$, $B^-$, $B^0$
and $\bar {B^0}$) mesons in the magnetized nuclear matter
are calculated using the chiral
effective model. The masses of these mesons within the model
are obtained from their interactions with the scalar mesons
and the nucleons. The number and scalar densities of the
proton, which is electrically charged,
have contributions from the Landau energy levels
in the presence of a magnetic field \cite{dmeson_mag,bmeson_mag}.
The calculations for the masses of the $B$
and $\bar B$ in the chiral effective model 
are performed using the mean field 
approximation, i.e., neglecting the effects of the Dirac sea. 
The magnetic field dependence of the masses of the open 
bottom mesons ($B$, $\bar B$, $B^*$ and $\bar {B^*}$) 
for $\rho_B=\rho_0$ for the isospin
symmetric ($\eta=0$) and asymmetric ($\eta$=0.5) 
nuclear matter, are shown in figures \ref{mbbar_PV_rh0_eta0}
and \ref{mbbar_PV_rh0_eta5}.
The masses of the open bottom vector mesons
($B^*$ and $\bar {B^*}$ mesons), which have the
same quark-antiquark constituents as $B$ and $\bar B$ mesons,
are assumed to have identical mass shifts as the shifts
in the masses of the $B$ and $\bar B$ mesons 
calculated within the chiral effective model.  
This is in line with the mass modifications of hadrons 
within the QMC model, which arise due to the modification
of the scalar density of the light quark (antiquark)
constituent of the hadron \cite{Hosaka_Prog_Part_Nucl_Phys}. 
To obtain the masses of the $B^*$ and $\bar {B^*}$ mesons
in the magnetized nuclear matter,
we assume, 
\begin{equation}
{m^*}_{{B^*}(\bar {B^*})}-m^{(vac)}_{{B^*}(\bar {B^*})}
={m^*}_{B(\bar B)}-m^{(vac)}_{B(\bar B)}
\label{mbstr}
\end{equation}
In figures \ref{mbbar_PV_rh0_eta0}
and \ref{mbbar_PV_rh0_eta5}, for $\rho_B=\rho_0$
and $\eta$=0 and 0.5,
the masses of the $B$ and $\bar B$ mesons,
as calculated using the chiral effective model
are shown as dotted lines. These
are observed to have marginal dependence 
on the magnetic field. This is due to the fact
that the scalar fields ($\sigma$, $\delta$, $\zeta$
and $\chi$) have negligible dependence on the magnetic field
for a given value of the baryon density and isospin
asymmetry parameter. In the mean field approximation,
we thus have the quark condensates, which are related
to the scalar fields as 
\begin{equation}
m_u\langle \bar u u\rangle =\frac{1}{2}{m_\pi}^2f_\pi 
(\sigma +\delta),\;\;
m_d\langle \bar d d\rangle =\frac{1}{2}{m_\pi}^2f_\pi 
(\sigma -\delta),
\end{equation}
to have marginal dependence on the magnetic field. 
In the presence of a magnetic field, the magnetic catalysis, 
which is the phenomenon of enhancement of the quark condensates
with increase in the magnetic field, arises from
the magnetic field dependent Dirac sea contribution
\cite{mag_catalysis}. In Ref. \cite{mag_catalysis}, 
within the Walecka model, in isospin symmetric nuclear matter
in the presence of a magnetic field, the mass of the nucleon 
(${M^*_N}= M_N-g_{N \sigma}\sigma)$ is also observed to be
substantially modified as a result of the magnetic catalysis,
due to the modification of the scalar field, $\sigma$ 
(which is proportional to the quark condensates).  
This effect is not observed 
in the present calculations which uses the mean field 
approximation, and, hence, neglects the Dirac sea 
contributions. The density effects 
on the quark condensates $(\sim (\sigma \pm \delta))$ and 
gluon condensates ($\sim \chi^4$) 
are the dominant medium effects, 
as compared to the effects from the isospin asymmetry
and the magnetic field of the medium 
\cite{dmeson_mag,bmeson_mag,charmonium_mag,vecqsr_mag}.

In the presence of the magnetic fields, there is mixing 
between the pseudoscalar meson and longitudinal component
of vector meson. The pseudoscalar meson-vectore meson (PV) 
mixing gives rise to modifications of their masses, 
with a drop (increase) in the mass of the pseudoscalar 
(longitudinal component of the vector) meson. 
The studies of the heavy quarkonium state in the presence 
of an external magnetic field within a QCD sum rule approach 
\cite{charmonium_mag_QSR,charmonium_mag_lee,Suzuki_Lee_2017} 
as well as solution of the Schr\"odingier 
equation for the heavy quarkonium bound state using an
an effective potential \cite{Alford_Strickland_2013} 
show that the mass modifications of these mesons due to PV 
mixing are important. 
In Refs. \cite{charmonium_mag_QSR,charmonium_mag_lee},
a phenomenological Lagrangian interaction 
${\cal L}_{PV\gamma}=({g_{PV}}/{m_{av}}) 
{e\tilde F_{\mu \nu}}(\partial^\mu P)V^\nu$
(with $m_{av}$ as the average of the masses of the
pseudoscalar and vector mesons),
was used for the charmonium states to study the PV 
($J/\psi-\eta_c$, $\psi'-\eta'_c$) mixing effects
in the presence of an external magnetic field. 
In the field theoretical model of composite hadrons
with quark (and antiquark) constituents as used
in the present work, 
the masses of the charmonium states have been studied
accounting for the PV ($J/\psi-\eta_c$, $\psi'-\eta'_c$
and $\psi(3770)-\eta'_c$) mixing effects 
in magnetized (nuclear) matter \cite{charmdw_mag}
using the above phenomenological Lagrangian interaction
\cite{charmonium_mag_QSR,charmonium_mag_lee,Suzuki_Lee_2017}.
The decay widths of $\psi (3770)\rightarrow D\bar D$ 
have been studied in Refs. \cite{charmdw_mag,open_charm_mag_AM_SPM},
which are observed to be modified significantly
due to the contributions of the $\psi(3770)-\eta'_c$ mixing
to the mass of the charmonium state $\psi(3770)$,
as well as due to the PV ($D-D^*$ and $\bar D-\bar {D^*}$) 
mixing contributions to the masses of the open charm mesons
\cite{open_charm_mag_AM_SPM}. The parameter $g_{PV}$
of the phenomenological Lagrangian density was 
calculated from the observed decay width of
$V\rightarrow P \gamma$ in vacuum. 
In the present study, we shall consider the PV mixing 
effects, in addition to the lowest Landau level contribution
to the masses of the charged open bottom mesons, 
for the study of the in-medium partial decay width
of $\Upsilon(4S)\rightarrow B\bar B$.
Due to lack of data on radiative processes 
($V\rightarrow P\gamma$) in the bottom sector,
we shall use the interaction Hamiltonian 
given by equation (\ref{H_spin_mixing})
to study the PV mixing effects on the masses 
of the open bottom mesons and $\Upsilon (4S)$ state,
and hence on the decay width for $\Upsilon (4S)
\rightarrow B \bar B$.

Strong magnetic fields are produced in non-central 
ultra-relativistic heavy ion collision experiments 
and the produced matter is extremely dilute. 
In the following, we shall consider the effects
of magnetic field on the masses of the open bottom
mesons and the $\Upsilon (4S)$ state for 
$\rho_B=0$ and $\rho_B=\rho_0$, and study their effects
on the decay width of $\Upsilon (4S)\rightarrow B\bar B$. 
We shall take into account the contributions 
due to the PV mixing effects 
(in addition to the LLL contributions
for the charged open bottom mesons)  
to the masses calculated in the chiral effective model
in the magnetized nuclear matter.

The masses of the open bottom (pseudoscalar
and vector) mesons are plotted for $\rho_B$=0 in figure 
\ref{mbbar_PV_rhbr0}. The effects of the lowest 
Landau level (LLL) for the $B^\pm$ as well as ${B^{*\pm}}^{(||)}$ 
mesons are observed to lead to an increase in their masses.
However, the effects due to the PV ($B-B^*$ and $\bar B-\bar {B^*}$)
mixings on the masses of these mesons 
are observed to dominate over LLL contributions.
These, along with modifications of the mass
of $\Upsilon(4S)$ (longitudinal component)
due to $\Upsilon(4S)^{||}-\eta_b(4S)$ mixing,
have significant effects on the decay widths 
of $\Upsilon (4S) \rightarrow B \bar B$ as can be seen from figure 
\ref{dwFT_upsilon_4s_etab4s_rh0}. For $\rho_B$=0,
the decay widths are plotted in (a) and (b).
In figure 7(a), the effects of PV mixing are not 
taken into account, whereas, these effects are considered
in 7(b). 
The vacuum mass of $\Upsilon(4S)$ is 10579.4 MeV.
The pseudoscalar meson $\eta_b(4S)$ is not yet observed
experimentally. We take the mass of $\eta_b(4S)$ to be
10573 MeV, as calculated from a relativistic potential
model \cite{Ebert}. 
When the PV mixing effects are not taken into consideration, 
the decay width of $\Upsilon(4S)$ to $B^+B^-$ is observed 
to decrease with increase in magnetic field and becomes zero 
at around $eB=4.5 m_ \pi ^ 2$ for $\rho_B=0$, as can be
seen from figure 7(a). 
This is due to the fact that
the masses of the charged $B^\pm$ increase due to the LLL 
contributions. On the other hand, the decay width 
for the neutral $B\bar B$ for $\rho_B$=0, remains same as 
its vacuum value, when the PV mixing effects are not considered,
as can be seen from figure 7(a). When the PV effetcs 
are taken into account, the decay width of $\Upsilon(4S)$
to $B^+B^-$ is observed initially to increase with increase 
in the magnetic field followed by a drop reaching a value
of around 21.8 MeV at around $eB/m_\pi^2\sim 6$. 
As the magnetic field is further increased, there is observed to be
a steep rise with the decay width reaching a value of 
around 177.6 (254) MeV at $eB/m_\pi^2=8.5(9)$. 
A similar behaviour for the decay width of 
$\Upsilon(4S)\rightarrow B^+B^-$ is observed for the case
when the $\Upsilon(4S)-\eta_b(4S)$
mixing contributions are not taken into account
(shown as dotted lines), however, still accounting for the mass
modifications of the open bottom mesons due to PV mixing.
For the decay width $\Upsilon(4S)\rightarrow B^0{\bar {B^0}}$
there is observed to be an increase with the magnetic field,
with the rise being steeper for $eB$ larger than  $8 m_\pi^2$.
The main difference between the two channels is due to 
the absence (presence) of the Landau level contributions for the
neutral (charged) $B\bar B$ mesons.
The behaviour of the decay width is determined by the 
value of the center of mass momentum, $|{\bf p}|$, which depends
on the masses of the $\Upsilon(4S)$, $B$ and $\bar B$ mesons,
through equation (\ref{pb}). The dependence of the decay widths
of $\Upsilon (4S)\rightarrow B\bar B (B^0{\bar B}^0, B^+B^-)$
are through a polynomial term multiplied by an exponential
term, as can be seen from the expression of the decay widths.
 
The in-medium masses of the $B$, $\bar B$, $B^*$ and $\bar B^*$
mesons at $\rho_B=\rho_0$  in symmetric and asymmetric nuclear 
matter in presence of magnetic field
are plotted as functions of $eB/m_\pi^2$
in figures \ref{mbbar_PV_rh0_eta0} and \ref{mbbar_PV_rh0_eta5} 
respectively. The mass of the ${B^*}^\pm$ due to the lowest 
Landau level contribution (n=0) is given as 
\begin{equation}
m^{eff}_{{B^*}^\pm}=\sqrt {{m^*_{{B^*}^\pm}}^2
+(-gS_z+1)|eB|},
\label{mbstrpm_landau_Sz}
\end{equation}
whereas the masses of the neutral ${B^*}^0$ and 
${\bar {B^*}}^0$  are given as
\begin{equation}
m^{eff}_{B^{*0},{{\bar B}^{*0}}}=
m^{*}_{B^{*0},{{\bar B}^{*0}}}.
\label{mbstr0bstr0bar}
\end{equation}
Equation (\ref{mbstrpm_landau_Sz}) refers to the mass of a
charged vector particle due to lowest Landau level,
ignoring its internal structure. As can be seen from
equation (\ref{mbstrpm_landau_Sz}), the mass depends on the
z-component of the spin, $S_z$. For $S_z=1$, the mass
squared of the vector particle decreases by $|eB|$, if we 
take the gyromagnetic ratio to be $g=2$. In the presence 
of a magnetic field, however, there is (PV) mixing between
the longitudinal component ($S_z=0$) 
of the $B^*({\bar B}^*)$ meson and
the pseudoscalar meson $B(\bar B)$, which gives
rise to mass modification of the $B(\bar B)$ meson.
The mass of the longitudinal component ($S_z$=0) 
of the charged $B^*({\bar B}^*)$ meson 
due to the lowest Landau level (LLL) contribution
is given by 
\begin{equation}
m^{eff}_{{{B^*}^\pm}^{(||)}}=\sqrt {{m^*_{{B^*}^\pm}}^2
+|eB|}.
\label{mbstrpm_landau}
\end{equation}
As has already been mentioned, the masses of the $B^*$
and $\bar {B^*}$ mesons in the nuclear medium,
occurring in equations (\ref{mbstr0bstr0bar}) and 
(\ref{mbstrpm_landau}) are obtained from the in-medium
masses of the $B(\bar B)$ mesons, ${m^*}_{B(\bar B)}$
using equation (\ref{mbstr}).
The LLL contributions lead to an increase
in the masses of the charged $B^\pm$ as well as of the
longitudinal component of the charged $(B^{*\pm})$ mesons.
As can be seen from figures \ref{mbbar_PV_rhbr0}, 
\ref{mbbar_PV_rh0_eta0} and \ref{mbbar_PV_rh0_eta5}, 
the $B-B^*$ and $\bar B-\bar {B^*}$ mixings lead to significant 
modifications to the masses of these mesons at $\rho_B$=0,
and at $\rho_B=\rho_0$ in the symmetric as well as
asymmetric nuclear matter. 

In figure \ref{mupsln_etab_4s}, the effects of the
PV mixing on the masses of 
the longitudinal component of $\Upsilon(4S)$ 
and the pseudoscalar meson $\eta_b(4S)$
are plotted as functions of $eB/(m_\pi^2)$.
These are shown for the symmetric and asymmetric
magnetized nuclear matter for $\rho_B=\rho_0$
and compared with the masses for $\rho_B=0$. 
The modifications of the mass of $\Upsilon(4S)^{(||)}$
due to PV mixing is observed to affect significantly
the decay width of $\Upsilon(4S)$ to $B^+B^-$ and 
$B^0\bar {B^0}$, as can be observed from figure 
\ref{dwFT_upsilon_4s_etab4s_rh0}.

The decay width of $\Upsilon(4S)$ to the charged
open bottom meson pair ( $B^+ B^-$), as can be seen
from figure \ref{dwFT_upsilon_4s_etab4s_rh0},
is observed to have an initial increase with magnetic field,
followed by a drop leading to vanishing of the decay
width and again a rise as the magnetic field
is further increased. The vanishing of the 
decay width (so called nodes) is similar to the
behaviour of the decay width with density,
for isospin symmetric matter
(see figures \ref{dwFT_amm_eb4} and \ref{dwFT_amm_eb10}).
The in-medium behaviour of the decay widths
of $\Upsilon(4S)\rightarrow B^+ B^-$ 
and $\Upsilon(4S)\rightarrow B^0 \bar {B^0}$, 
are determined through the center 
of mass momentum, $|{\bf p}|$. These decay widths,
given by equations (\ref{gammaupslnbpbm}) and
(\ref{gammaupslnb0b0b}), 
have a polynomial term multiplied by an exponential term, 
dependence on $|{\bf p}|$ (given by equation (\ref{pb})) 
which is expressed in terms of the in-medium masses
of the $\Upsilon (4S)$, $B$ and $\bar B$ mesons.
The suppression of the decay channel to $B^+B^-$
as compared to the $B^0\bar {B^0}$ should show
as the charged open bottom mesons to be suppressed
as compared to the neutral $B\bar B$ mesons.

The modification of the mass of $\Upsilon(4S)$, due to
mixing with $\eta_b(4S)$ is observed to be quite
appreciable.
The mass of $\eta_b(4S)$ is calculated 
using equation (\ref{masspsi}), from the change 
in the dilaton field within the chiral effective model. 
The values of the masses (in MeV) of $\Upsilon(4S)$ and 
$\eta_b(4S)$ are obtained to be 
10535.69 and 10531.22 at $\rho_B=\rho_0$
in isospin symmetric ($\eta$=0) and
10542.66  and 10537.88 in asymmetric 
($\eta$=0.5) nuclear matter for $B=0$.
The modifications of these masses are negligible
as the magnetic field is increased,
when the PV mixing is not taken into account.
Considering the PV ($\Upsilon (4S)-\eta_b(4S)$) mixing,
at $\rho_B=\rho_0$, for $eB=10m_\pi^2$,
the mass of the $\Upsilon(4S)$ 
($\eta_b(4S)$) is obtained as
10545.9 (10521.46) MeV for isospin 
symmetric nuclear matter ($\eta$=0), 
and 10552.5 (10528) MeV 
in asymmetric nuclear matter (with $\eta=0.5$).
The appreciable contribution due to the mass modification
of $\Upsilon(4S)$ from PV mixing, can be seen from
the dotted lines in (b), (d)
and (f) of figure \ref{dwFT_upsilon_4s_etab4s_rh0},
which correspond to the cases when PV mixing 
is not taken into account for the mass of
$\Upsilon(4S)$, which are observed to be significantly
modified when the PV mixing 
contributions for $\Upsilon (4S)$ are taken into account. 
The PV mixing effects of the open bottom mesons
as well as of the $\Upsilon(4S)$ state
on their masses are thus the most important effects
due to the presence of the magnetic field
for the study of the in-medium decay width
of $\Upsilon (4S)\rightarrow B \bar B$
in the magnetized (nuclear) matter.

\section{Summary}
In the present work, we have used a composite model for hadrons 
with quark (and antiquark) constituents, to calculate the
in-medium decay widths of $\Upsilon \rightarrow B\bar B$
in nuclear matter in the presence of strong magnetic fields.
The effects of the isospin
asymmetry are observed to be large at high densities,
and more prominent for higher values of the magnetic
fields. In the presence of strong magnetic fields,
the decay widths of $\Upsilon (4S)$ to the charged
$B\bar B$ are observed to be suppressed as compared
to the decay to neutral $B^0 \bar {B^0}$ pair,
when the PV mixing effects are not taken into
consideration. This is due to the positive 
mass shifts of the charged $B^+$ and $B^-$ 
mesons arising from the lowest Landau level
contributions.
The PV ($B -B^*$ and ${\bar B}-{\bar {B^*}}$) 
mixing effects are observed to lead to 
appreciable modifications 
to the masses of the $B$ and $\bar B$ 
mesons, which, in turn, are observed to have dominant
modifications to the partial decay widths
of $\Upsilon(4S)\rightarrow B\bar B$.
The mass shift of $\Upsilon(4S)$ due to
mixing with $\eta_b(4S)$ is observed to be 
quite appreciable and has significant effect 
on the decay width. 
The significant modifications of the decay widths
of $\Upsilon(4S)$ to the charged and neutral $B\bar B$ 
should show in the production of these
mesons and $\Upsilon(4S)$ in peripheral 
ultra-relativistic heavy ion collision 
experiments, e.g. at RHIC and LHC,
where the produced magnetic field is huge.

\begin{section}*{Acknowledgements}
Amruta Mishra acknowledges financial support
from Department of Science and Technology (DST),
Government of India (project no. CRG/2018/002226).
\end{section}



\begin{thebibliography}{}
\bibitem{Hosaka_Prog_Part_Nucl_Phys} 
A. Hosaka, T. Hyodo, K. Sudoh, Y. Yamaguchi, S. Yasui,
Prog. Part. Nucl. Phys. {\bf 96}, 88 (2017).
\bibitem{HIC_mag_1} V. Skokov, A. Y. Illarionov and V. Toneev,
Int. J. Mod. Phys. A {bf 24}, 5925 (2009).
\bibitem{HIC_mag_2}
 W. T. Deng and X.G.Huang, Phys.Rev. C {\bf 85}, 044907 (2012).
\bibitem{HIC_mag_3}
 D. Kharzeev, L. McLerran and 
H. Warringa, Nucl. Phys. A {\bf 803}, 227 (2008).
\bibitem{HIC_mag_4}
 K. Fukushima, D. E. Kharzeev and H. J. Warringa, 
Phys. Rev. D {\bf 78}, 074033 (2008).

\bibitem{Gubler_D_mag_QSR} P. Gubler, K. Hattori, S. H. Lee, M. Oka,
S. Ozaki and K. Suzuki, Phys. Rev. D {\bf 93}, 054026 (2016).

\bibitem{machado_1} C. S. Machado, F. S. Navarra. E. G. de Oliveira
and J. Noronha, Phys. Rev. D {\bf 88}, 034009 (2013).

\bibitem{B_mag_QSR} C. S. Machado, R.D. Matheus, S.I. Finazzo and
J. Noronha, Phys. Rev. D {\bf 89}, 074027 (2014).

\bibitem{dmeson_mag}
Sushruth Reddy P, Amal Jahan CS, Nikhil Dhale, Amruta Mishra,
J. Schaffner-Bielich, Phys. Rev. C {\bf 97}, 065208 (2018).
\bibitem{bmeson_mag}
Nikhil Dhale, Sushruth Reddy P, Amal Jahan CS, Amruta Mishra,
Phys. Rev. C {\bf 98}, 015202 (2018).

\bibitem{charmonium_mag}
Amal Jahan CS, Nikhil Dhale, Sushruth Reddy P,
Shivam Kesarwani, Amruta Mishra, 
Phys. Rev. C {\bf 98}, 065202 (2018).

\bibitem{upsilon_mag}
Amal Jahan CS, Shivam Kesarwani, Sushruth Reddy P,
Nikhil Dhale, Amruta Mishra, arXiv: 1807.07572 (nucl-th).

\bibitem{charmonium_mag_QSR} S. Cho, K. Hattori, S. H. Lee, K. Morita
and S. Ozaki, Phys. Rev. Lett. {\bf 113}, 122301 (2014).
\bibitem{charmonium_mag_lee} 
S. Cho, K. Hattori, S. H. Lee, K. Morita
and S. Ozaki, Phys. Rev. D {\bf 91}, 045025 (2015).

\bibitem{Suzuki_Lee_2017}
K. Suzuki and S. H. Lee, Phys. Rev. C {\bf 96}, 035203 (2017).

\bibitem{Alford_Strickland_2013}
J. Alford and M. Strickland, Phys. Rev. D {\bf 88}, 105017
(2013).


\bibitem{time_evolution_B_HIC_Tuchin_1}
K. Tuchin, Phys. Rev. C {\bf 83}, 017901 (2011).
\bibitem{time_evolution_B_HIC_Tuchin_2}
K. Marasinghe and K. Tuchin, Phys. Rev. C {\bf 84},
044908 (2011).
\bibitem{time_evolution_B_HIC_Tuchin_3}
K. Tuchin, Phys. Rev. C {\bf 83}, 034904 (2010),
K. Tuchin, Erratum Phys. Rev. C {\bf 83}, 039903(E) (2011).
\bibitem{time_evolution_B_HIC_Tuchin_4}
K. Tuchin, Phys. Rev. C {\bf 88}, 024911 (2013).

\bibitem{time_evolution_B_HIC_Ajit}
Arpan Das, S. S. Dave. P.S. Saumia and A.M. Srivastava, 
Phys. Rev. C {\bf 96}, 034902 (2017).

\bibitem{kimlee}
Sugsik Kim, Su Houng Lee, Nucl. Phys. A {\bf 679}, 517 (2001).
\bibitem{klingl} F. Klingl, S. Kim, S. H. Lee, P. Morath and
W. Weise, Phys. Rev. Lett. {\bf 82}, 3396 (1999).
\bibitem{amarvjpsi_qsr} 
Arvind Kumar and Amruta Mishra, Phys. Rev. C {\bf 82}, 045207 (2010).
\bibitem{jpsi_etac_mag}
Pallabi Parui, Ankit Kumar, Sourodeep De, Amruta Mishra,
arXiv: 1811.04622 (nucl-th).
\bibitem{moritalee_1}
K. Morita and S.H. Lee, Phys. Rev. C {\bf 77}, 064904 (2008).
\bibitem{moritalee_2}
S.H. Lee and K. Morita, Phys. Rev. D {\bf 79}, 011501(R) (2009).
\bibitem{moritalee_3}
K. Morita and S.H. Lee, Phys. Rev. C {\bf 85}, 044917 (2012).
\bibitem{moritalee_4}
K. Morita and S.H. Lee, Phys. Rev. Lett {\bf 100}, 022301 (2008).

\bibitem{open_heavy_flavour_qsr_1} 
Arata Hayashigaki , Phys. Lett. B {\bf 487}, 96 (2000).
\bibitem{open_heavy_flavour_qsr_2} 
T. Hilger, R. Thomas and B. K\"ampfer, Phys. Rev. C {bf 79},
025202 (2009).
\bibitem{open_heavy_flavour_qsr_3} 
 T. Hilger, B. K\"ampfer and S. Leupold, 
Phys. Rev. C {\bf 84}, 045202 (2011). 
\bibitem{open_heavy_flavour_qsr_4} 
S. Zschocke, T. Hilger and B. K\"ampfer,
Eur. Phys. J. A {\bf 47} 151 (2011).

\bibitem{Wang_heavy_mesons_1} 
Z-G. Wang and Tao Huang, Phys. Rev. C {\bf 84}, 048201 (2011).
\bibitem{Wang_heavy_mesons_2} 
Z-G. Wang, Phys. Rev. C {\bf 92}, 065205 (2015).

\bibitem{arvind_heavy_mesons_QSR_1}
Rahul Chhabra and Arvind Kumar, 
Eur. Phys. J A {\bf 53}, 105 (2017).
\bibitem{arvind_heavy_mesons_QSR_2}
Rahul Chhabra and Arvind Kumar, 
Eur. Phys. J C {\bf 77}, 726 (2017). 
\bibitem{arvind_heavy_mesons_QSR_3}
Arvind Kumar and Rahul Chhabra, 
Phys. Rev. C {\bf 92}, 035208 (2015).


\bibitem{eichten_1}
E.Eichten, K. Gottfried, T. Kinoshita, K.D. Lane and T.M. Yan, 
Phys. Rev. D {\bf 17}, 3090 (1978).
\bibitem{eichten_2}
E.Eichten, K. Gottfried, T. Kinoshita, K.D. Lane and T.M. Yan, 
Phys. Rev. D {\bf 21}, 203 (1980).
\bibitem{satz_1}
L. Kluberg and H. Satz, arXiv:0901.3831 (hep-ph).
\bibitem{satz_2}
F. Karsch, M. T. Mehr, and H. Satz, 
Z. Phys. C {\bf 37}, 617 (1988).
\bibitem{satz_3}
A. Bazavov, P. Petreczky, and A. Velytsky, 
arXiv: 0904.1748 (hep-ph).
\bibitem{satz_4}
S. Digal, P. Petreczky, and H. Satz, 
Phys. Lett. B {\bf 514}, 57 (2001). 
\bibitem{satz_5}
A. Mocsy and P. Petreczky, 
Phys. Rev. D {\bf 73}, 074007 (2006).
\bibitem{repko}
S.F. Radford and W W. Repko, Phys.Rev D {\bf 75}, 074031 (2007).

\bibitem{Ebert}
D. Ebert, R. N. Faustov, V. O. Galkin, Phys. At. Nuclei {\bf 76}, 
1554 (2013); {\em ibid}, Eur. Phys. Jour. C {\bf  71}, 1825 
(2011).

\bibitem{Bonati_pot_model}
C. Bonati, M. D. Elia, A. Rucci, Phys. Rev. D {\bf 92},
054014 (2015).

\bibitem{Yoshida_Suzuki_heavy_flavour_meson_strong_B}
T. Yoshida and K. Suzuki, Phys. Rev. D {\bf 94}, 074043
(2016).

\bibitem{ltolos} L.Tolos, J. Schaffner-Bielich and A. Mishra,
Phys. Rev. {\bf C 70}, 025203 (2004).
\bibitem{ljhs} L. Tolos, J. Schaffner-Bielich
and H. St\"ocker, Phys. Lett. {\bf B 635}, 85 (2006).
\bibitem{mizutani_1} 
T. Mizutani and A. Ramos, Phys. Rev. {\bf C 74},
065201 (2006).
\bibitem{mizutani_2} 
 L.Tolos, A. Ramos and T. Mizutani, Phys. Rev. {\bf C 77},
015207 (2008).
\bibitem{HL}J. Hofmann and M.F.M.Lutz, 
Nucl. Phys. {\bf A 763}, 90 (2005).

\bibitem{tolos_heavy_mesons_1}
R. Molina, D. Gamermann, E. Oset, and L. Tolos,
Eur. Phys. J A {\bf 42}, 31 (2009).
\bibitem{tolos_heavy_mesons_2}
L. Tolos, R. Molina, D. Gamermann, and E. Oset, 
Nucl. Phys. A {\bf 827} 249c (2009).

\bibitem{open_heavy_flavour_qmc_1}
	K. Tsushima, D. H. Lu, A. W. Thomas, K. Saito, and R. H. Landau,
	Phys. Rev. C {\bf 59}, 2824 (1999).
\bibitem{open_heavy_flavour_qmc_2}
	A. Sibirtsev,	K. Tsushima, and A. W. Thomas,
	Eur. Phys. J. A {\bf 6}, 351 (1999).
\bibitem{open_heavy_flavour_qmc_3}
        K. Tsushima and F. C. Khanna, Phys. Lett. B {\bf 552}, 138
        (2003).
\bibitem{qmc_1} 
P. A. M. Guichon, Phys. Lett. B {\bf 200}, 235 (1988).
\bibitem{qmc_2} 
 K. Saito and A. W. Thomas, Phys. Lett B {\bf 327}, 9 (1994).
\bibitem{qmc_3} 
K. Saito, K. Tsushima and A. W. Thomas, Nucl. Phys. A {\bf 609}, 339
(1996).
\bibitem{qmc_4} 
 P.K. Panda, A. Mishra, J. M. Eisenberg and W. Greiner,
Phys. Rev. C {\bf 56}, 3134 (1997).
\bibitem{krein_jpsi}
G. Krein, A. W. Thomas and K. Tsushima, Phys. Lett. B {\bf 697},
136 (2011).
\bibitem{krein_17}
G. Krein, A. W. Thomas and K. Tsushima, arXiv: 1706.02688
(hep-ph).

\bibitem{Yasui_Sudoh_pion}
S. Yasui and K. Sudoh, Phys. Rev. C {\bf 87}, 015202 (2013).
\bibitem{Yasui_Sudoh_heavy_meson_Eff_th}
S. Yasui and K. Sudoh, Phys. Rev. C {\bf 89}, 015201 (2014).

\bibitem{Yasui_Sudoh_heavy_particle_impurity}
S. Yasui and K. Sudoh, Phys. Rev. C {\bf 88}, 015201 (2013).


\bibitem{leeko} Su Houng Lee and Che Ming Ko, 
Phys. Rev. C {\bf 67}, 038202 (2003).
\bibitem{pes1} M.E. Peskin, Nucl. Phys. {\bf B156}, 365 (1979).
\bibitem{pes2} 
G. Bhanot and M.E. Peskin, Nucl. Phys. {\bf B156}, 391 (1979).
\bibitem{voloshin}
M.B.Voloshin, Nucl. Phys. B154 ,365 (1979).


\bibitem{Schechter} J. Schechter, Phys. Rev. D {\bf 21}, 3393 (1980).
\bibitem{paper3}
 	P. Papazoglou, D. Zschiesche, S. Schramm, J. Schaffner-Bielich,
	H. St\"ocker, and W. Greiner, Phys. Rev. C {\bf 59},  411  (1999).
\bibitem{kristof1}
	A. Mishra, K. Balazs, D. Zschiesche, S. Schramm,
	H. St\"ocker, and W. Greiner,
        Phys. Rev. C {\bf 69}, 024903 (2004). 
\bibitem{amarvdmesonTprc}
Arvind Kumar and Amruta Mishra, Phys. Rev. C {\bf 81}, 065204
(2010).
\bibitem{amarvepja}
Arvind Kumar and Amruta Mishra, Eur. Phys. A {\bf 47}, 164
(2011).
\bibitem{AM_DP_upsilon}
Amruta Mishra and Divakar Pathak, Phys. Rev. C {\bf 90}, 025201
(2014).

\bibitem {amdmeson} 
A. Mishra, E. L. Bratkovskaya, J. Schaffner-Bielich, 
S.Schramm and H. St\"ocker, Phys. Rev. {\bf C 69}, 015202 (2004).
\bibitem{amarindamprc}
Amruta Mishra and Arindam Mazumdar, Phys. Rev. C {\bf 79},  024908 (2009). 
\bibitem{DP_AM_Ds}
Divakar Pathak and Amruta Mishra, Adv. High Energy Phys. 2015,
697514 (2015).
\bibitem{DP_AM_bbar}
Divakar Pathak and Amruta Mishra, Phys. Rev. C {\bf 91}, 045206
(2015).
\bibitem{DP_AM_Bs}
Divakar Pathak and Amruta Mishra, Int. J. Mod. Phy. E {\bf 23}, 
1450073 (2014).


\bibitem{hartree}
	D. Zschiesche, A. Mishra, S. Schramm, H. St\"ocker and W. Greiner,
        Phys. Rev. C {\bf 70}, 045202 (2004).
\bibitem{kaon_antikaon}
A. Mishra, E. L. Bratkovskaya, J. Schaffner-Bielich, S. Schramm
     and H. St\"ocker, Phys. Rev. C {\bf 70}, 044904 (2004).
\bibitem{isoamss}
A. Mishra and S. Schramm, Phys. Rev. C {\bf 74}, 064904 (2006).	
\bibitem{isoamss1}
A. Mishra, S. Schramm and W. Greiner, Phys. Rev. C {\bf 78}, 
024901 (2008).
\bibitem{isoamss2}
Amruta Mishra, Arvind Kumar, Sambuddha Sanyal, S. Schramm,
Eur. Phys, J. A {\bf 41}, 205  (2009).  
\bibitem{pneutronstar}
Amruta Mishra, Arvind Kumar, Sambuddha Sanyal, V. Dexheimer, 
S. Schramm, Eur. Phys. J {\bf 45}, 169 (2010).
\bibitem{am_vecmeson_qsr} Amruta Mishra, 
Phys. Rev. C {\bf 91} 035201 (2015).
\bibitem{vecqsr_mag}
Amruta Mishra, Ankit Kumar, Pallabi Parui, Sourodeep De,
Phys. Rev. C {\bf 100} 015207 (2019).

\bibitem{kmeson_mag}
Amruta Mishra, Anuj Kumar Singh, Neerajj Singh Rawat, Pratik Aman,
Eur. Phys. Jour. A 55, 107 (2019).

\bibitem{friman}
B. Friman, S. H. Lee and T. Song, Phys. Lett, B {\bf 548}, 153
(2002).

\bibitem{3p0_1}
A. Le Yaouanc, L. Oliver, O. Pene and  J. C. Raynal, 
Phys. Rev. D {\bf 8}, 2223 (1973).
\bibitem{3p0_2}
A. Le Yaouanc, L. Oliver, O. Pene and  J. C. Raynal, 
Phys. Rev. D {\bf 9}, 1415 (1974).
\bibitem{3p0_3}
A. Le Yaouanc, L. Oliver, O. Pene and  J. C. Raynal, 
ibid, Phys. Rev. D {\bf 11}, 1272 (1975).
\bibitem{3p0_4}
T. Barnes, F. E. Close, P. R. Page and E. S. Swanson, Phys. Rev. D
{\bf 55}, 4157 (1997).

\bibitem{amspmwg}
Amruta Mishra, S. P. Misra and W. Greiner, Int. J. Mod. Phys.
E {\bf 24}, 155053 (2015).
\bibitem{amspm_upsilon}
Amruta Mishra and S. P. Misra, Phy. Rev. C {\bf 95}, 065206 (2017).

\bibitem{charmdecay_mag}
 A. Mishra , A. Jahan CS , S. Kesarwani , H. Raval , S. Kumar, and J.
Meena , Eur. Phys. J. A 55,99 (2019).

\bibitem{charmdw_mag}
 Amruta Mishra, S.P. Misra,
Phys. Rev. C {\bf 102}, 045204 (2020).

\bibitem{open_charm_mag_AM_SPM}
Amruta Mishra and S. P. Misra, Int. Jour. Mod. Phys. E {\bf 30},
2150064 (2021).

\bibitem{strange_AM_SPM}
Amruta Mishra and S. P. Misra, Int. Jour. Mod. Phys. E {\bf 30},
2150014 (2021).

\bibitem{Quarkonia_B_Iwasaki_Oka_Suzuki}
S. Iwasaki, M. Oka, K. Suzuki, Eur. Phys. Jour. A {\bf 57}
(2021) 222.
\bibitem{Chernodub}
M.N. Chernodub, Lect. Notes Phys. {\bf 871}, 143 (2013);
M. N. Chernodub, Phys. Rev. D {\bf 82}, 085011 (2010).

\bibitem{Taya}
H. Taya, Phys. Rev. D {\bf 92}, 014038 (2015).

\bibitem{spm781} S. P. Misra, Phys. Rev. D {\bf 18}, 1661 (1978).
\bibitem{spm782} S. P. Misra, Phys. Rev. D {\bf 18}, 1673 (1978).
\bibitem{spmdiffscat} S. P. Misra and L. Maharana, Phys. Rev. D
{\bf 18}, 4103 (1978).

\bibitem{broderick1}
A. Broderick, M. Prakash and J.M.Lattimer, Astrophys. J. 537, 351
(2002). 
\bibitem{broderick2}
A.E. Broderick, M. Prakash and J. M. Lattimer, Phys. Lett. {\bf B}
531, 167 (2002).
\bibitem{Wei}
F. X. Wei, G. J. Mao, C. M. Ko, L. S. Kisslinger, H. St\"ocker,
and W. Greiner, J. Phys. G, Nucl. Part. Phys. {\bf 32},
47 (2006).
\bibitem{mao} Guang-Jun Mao, Akira Iwamoto, Zhu-Xia Li,
Chin. J. Astrophys. 3, 359 (2003).
\bibitem{amm}
M. Pitschmann and A. N. Ivanov, arXive : 1205.5501 (math-ph). 

\bibitem{VD_SS_1}
V. Dexheimer, R. Negreiros, S. Schramm, Eur. Phys. Journal A 
{\bf 48}, 189 (2012).
\bibitem{VD_SS_2}
 V. Dexheimer, B. Franzon and S. Schramm,
Jour. Phys. Conf. Ser. {\bf 861}, 012012 (2017).
\bibitem{aguirre_fermion}
R. M. Aguirre and A. L. De Paoli, Eur. Phys. J. A {\bf 52}, 343
(2016).
\bibitem{mag_catalysis}  A. Haber, F. Preis and A. Schmitt,
Phys. Rev. D {\bf 90}, 125036 (2014).


\end{thebibliography}
\end{document}